\documentclass[prb,twocolumn,floatfix,reprint,aps,superscriptaddress]{revtex4}

\usepackage{amsfonts}
\usepackage{mathrsfs}
\usepackage{amsmath}
\usepackage{color}
\usepackage{graphicx}
\usepackage{bm}
\usepackage{amssymb}
\usepackage{xspace}
\usepackage{epstopdf}
\usepackage{dcolumn}
\usepackage{longtable}
\usepackage{multirow}
\usepackage[colorlinks=true, letterpaper=true, pdfstartview=FitV, linkcolor=blue, citecolor=blue, urlcolor=blue]{hyperref}

\usepackage{ulem}

\begin{document}

\title{Orientation-dependent Rashba spin-orbit coupling of two-dimensional hole gases in semiconductor quantum wells: linear or cubic }

\author{Jia-Xin Xiong}
\affiliation{State Key Laboratory of Superlattices and Microstructures, Institute of Semiconductors,Chinese Academy of Sciences, Beijing 100083, China}
\affiliation{Center of Materials Science and Optoelectronics Engineering, University of Chinese Academy of Sciences, Beijing 100049, China}

\author{Shan Guan}
\email{shan\_guan@semi.ac.cn}
\affiliation{State Key Laboratory of Superlattices and Microstructures, Institute of Semiconductors,Chinese Academy of Sciences, Beijing 100083, China}

\author{Jun-Wei Luo}
\email{jwluo@semi.ac.cn}
\affiliation{State Key Laboratory of Superlattices and Microstructures, Institute of Semiconductors,Chinese Academy of Sciences, Beijing 100083, China}
\affiliation{Center of Materials Science and Optoelectronics Engineering, University of Chinese Academy of Sciences, Beijing 100049, China}
\affiliation{Beijing Academy of Quantum Information Sciences, Beijing 100193, China}

\author{Shu-Shen Li}
\affiliation{State Key Laboratory of Superlattices and Microstructures, Institute of Semiconductors,Chinese Academy of Sciences, Beijing 100083, China}
\affiliation{Center of Materials Science and Optoelectronics Engineering, University of Chinese Academy of Sciences, Beijing 100049, China}

\begin{abstract}
Rashba spin-orbit coupling (SOC) described by odd powers of $\bf{k}$ terms in semiconductor quantum wells (QWs) plays a critical role in spintronics and quantum computing. It was believed that the Rashba SOC in two-dimensional hole gases (2DHGs) is a $\bf{k}$-cubic term as the lowest order, in sharp contrast to its electron counterpart. Our recent work [Phys. Rev. B 103, 085309 (2021)] uncovered the emergence of $\bf{k}$-linear Rashba SOC in 2DHGs arising from their direct dipolar coupling to the external electric field in the presence of heavy-hole-light-hole (HH-LH) mixing. Here, we explore this $\bf{k}$-linear Rashba SOC for 2DHGs in Ge/Si QWs with different orientations and find the underlying physical mechanism of the emergence and strength of both linear and purely cubic Rashba SOC. By performing atomistic pseudopotential method calculations associated with symmetry analysis, we demonstrate that the upper bound of the $\bf{k}$-linear Rashba SOC of 2DHGs is in the [110]-oriented QWs, and that the purely cubic Rashba SOC without $\bf{k}$-linear term occurs in the [111]-oriented QWs with an even number of the well monolayers (MLs). The odevity of the well MLs can change the symmetry of the [001]- and [111]-oriented QWs and thus the behavior of spin splittings, providing direct evidence of the hypothesis that the HH-LH mixing nominally forbidden by the global symmetry $D_{2d}$ in [001]-oriented GaAs/AlAs QWs results from their local interface symmetry $C_{2v}$. Moreover, we illustrate that the $\bf{k}$-linear Rashba term in [001]-oriented QWs and odd-ML [111]-oriented QWs completely arises from the local interface induced HH-LH mixing, whereas the linear Rashba SOC in other orientations dominantly originates from the intrinsic HH-LH mixing. We further assess the intrinsic HH-LH mixing to illuminate its dependence to the $\bf{k}$-linear Rashba term. Our findings clarify the physical mechanism underlying the orientation-dependence of the Rashba SOC and provide design rules to realize $\bf{k}$-linear or purely $\bf{k}$-cubic Rashba SOC in 2D hole systems.
\end{abstract}


\maketitle

\section{Introduction}\label{section1}

The relativistic spin-orbit interaction couples the electron spin to its orbital motion in solids lacking spatial inversion symmetry due to electrons moving in an asymmetric electric field encountering an effective magnetic field in their frame of motion that couples to the electron's spin magnetic moment~\cite{Dresselhaus1955, Rashba1984, Meier2007}. There are two types of spin-orbit coupling (SOC) effects. One originates from bulk-inversion asymmetry (BIA) induced asymmetric crystal field, known as the Dresselhaus effect~\cite{Dresselhaus1955}. Another type arises from the potential-asymmetry induced electric field associated with structural-inversion asymmetry (SIA) in heterostructures, known as the Rashba effect~\cite{Rashba1984}. It has now been extended to bulk crystals due to intrinsic dipole field~\cite{Ishizaka_NM2011, LuoNP2014}. The Dresselhaus and Rashba SOC effects enable a wide variety of fascinating phenomena and have inspired a vast number of innovative concepts far beyond semiconductors, rendering them to play a crucial role in diverse fields of condensed matter physics~\cite{Manchon2015}.

In contrast to the Dresselhaus SOC, the Rashba SOC is more compelling due to its electrical tunability. It renders low-dimensional semiconductors becoming platforms promising for an increasing number of physical effects and device applications, such as spin Hall effect~\cite{Sinova2004, Bernevig2005, Awschalom2004}, spin galvanic effect~\cite{Wunderlich2009, Ivchenko1978, Ganichev2008}, spin transistors~\cite{Datta1990, Schliemann2003}, and spin qubits~\cite{Maurand2016, Watzinger2018}. In these physical effects and potential applications, long spin lifetimes are crucial. Regarding holes having a much longer spin lifetime than electrons due to the suppression of the contact hyperfine interaction between the nuclear spins and carrier spins~\cite{Watzinger2018, Maurand2016, Hendrickx2018, Brauns2016}, a large fraction of research attention has recently been shifted towards hole systems~\cite{Watzinger2018, Maurand2016, Hendrickx2018, Brauns2016}. Unlike group III-V semiconductors, group IV Ge and Si contain only less than 5\% of atomic nuclei with non-zero spins, which can be further engineered by isotopic purification~\cite{Tyryshkin2012, Veldhorst2014, Yoneda2018} into (almost) nuclear-spin-free materials, and thus can obtain the longest spin lifetimes among all semiconductors~\cite{Hu2011, Warburton2013}. Furthermore, Ge has the highest hole mobility among all known semiconductors and is compatible with mature Si Complementary-Metal-Oxide-Semiconductor (CMOS) technology. Moreover, Ge possesses a much stronger SOC than Si because strong spin-orbit interaction is an inherent relativistic effect of heavy atoms. All these factors jointly facilitate the replacement of Si by Ge towards scalable quantum computing. Specifically, Veldhorst and his co-workers have very recently implemented fast single-qubit~\cite{Hendrickx2018}, two-qubit~\cite{Hendrickx2020} and four-qubit logic operations~\cite{Hendrickx2021} using hole spins confined in planar Ge quantum dots (QDs) formed by top-gated Ge/SiGe quantum wells (QWs)~\cite{Sammak2021}, where the achieved fast qubit manipulation gives the credit to the strong SOC~\cite{Hendrickx2020, EDSR2007, Bulaev2005}.

Although the expected strong SOC renders Ge planar hole QDs central to the quest for electrical spin qubit manipulation enabling fast, low-power, and scalable quantum computation~\cite{Hendrickx2020, Scappucci2020}, the underlying microscopic physics remains unclear. In the material platform of 2D Ge/SiGe QWs, holes are exclusively located in heavy-hole (HH) subbands since the light-hole (LH) subbands are lifted by the strong quantum confinement effect, resulting in a large HH-LH splitting with a magnitude of about 100 meV~\cite{Scappucci2020}. The ground HH subband in semiconductor QWs was commonly believed to have vanishing $\bf{k}$-linear terms and instead $\bf{k}$-cubic terms as the lowest order in their Rashba SOC Hamiltonian~\cite{Luo2011, Xiong_PRB2021, Moriya2014, Winkler2003, Kloeffel2018a}. It is in sharp contrast to the electron counterpart that possesses a $\bf{k}$-linear Rashba SOC.  The $\bf{k}$-linear term, if present, tends to overwhelm all higher-order terms in the SOC Hamiltonian because the orbital motion of the free charge carriers is in a small-$\bf{k}$ range ($k_F\ll 1$). The expectation of 2D holes has a stronger SOC than 2D electrons is thus questionable~\cite{Hendrickx2020}. This puzzle has now been resolved by our recent discovery of finite $\bf{k}$-linear Rashba SOC in two-dimensional hole gases (2DHGs). It was found to be originating from a direct dipolar coupling of HH subbands to the external electric field in the presence of HH-LH mixing~\cite{Xiong_PRB2021}. We found that the strength of the $\bf{k}$-linear Rashba SOC depends highly on the growth direction of QWs. In [001]-oriented Ge/Si QWs, the $\bf{k}$-linear Rashba parameter $\alpha_R$ is less than 5 meV{\AA} in a modest range of applied electric field ($<100$ kV/cm). Whereas, in [110]-oriented QWs, $\alpha_R$ can exceed 90 meV{\AA}~\cite{Xiong_PRB2021}. The quest for strong Rashba SOC to achieve scalable quantum computing~\cite{Hendrickx2020, Scappucci2020} stimulates us to find the strongest $\bf{k}$-linear Rashba SOC of 2DHGs reachable in Ge/Si QWs by varying the growth direction.

On the other hand, the previously believed purely $\bf{k}$-cubic Rashba SOC in 2DHGs has already inspired the exploration of their unique benefits to spintronics~\cite{Schliemann2005, Moriya2014, Failla_PRB2015}. The cubic term has striking differences from the linear term in the effective magnetic field and the resulting spin-momentum locking-induced spin texture~\cite{Nakamura2012}. It has become an interesting subject and the search of $\bf{k}$-cubic Rashba SOC has even been extended to SrTiO3 surface~\cite{Nakamura2012, Shanavas2016},  SrTiO3-based oxide heterstructures~\cite{Lin2019}, surfaces~\cite{Usachov_PRL2020}, quantum point contacts~\cite{Chesi_PRL2011}, as well as to the discovery of new classes of bulk materials~\cite{Zhao_PRL2020}. However, the emergence of finite $\bf{k}$-linear Rashba SOC excludes semiconductor 2DHGs to be platforms for Rashba physics based on purely $\bf{k}$-cubic Rashba SOC. The mature Si microelectronic technology facilities us to seek also purely $\bf{k}$-cubic SOC in 2DHGs by varying growth directions of Ge/Si QWs.

In this work, we theoretically investigate the dependence of the Rashba SOC of 2DHGs in Ge/Si QWs on crystal orientations by using the atomistic semi-empirical pseudopotential method (SEPM)~\cite{Wang1994, Wang1995, Wang1999}. We find the upper bound of the linear Rashba SOC in the [110]-oriented Ge/Si QWs and purely $\bf{k}$-cubic Rashba SOC in the [111]-oriented Ge/Si QWs with an even number of the well monolayers (MLs). We further conduct the symmetry analysis and point out that the absence of $\bf{k}$-linear Rashba SOC in even-ML [111]-oriented QWs results from the prohibition of HH-LH mixing as enforced by the symmetry of both global crystal and local interface. The relatively weak linear Rashba SOC in the [001]-oriented QWs and odd-ML [111]-oriented QWs arises from the local-interface-induced HH-LH mixing. By doing the transformation of the Luttinger-Kohn (LK) Hamiltonian, we illustrate that the strongest linear Rashba SOC occurred in the [110]-oriented QWs among different growth directions is due to their strongest intrinsic HH-LH mixing induced by the QW confinement, which might be dominant over the local interface for HH-LH mixing.

The rest of this paper is organized as follows. Sec.~\ref{section2} introduces briefly the computational methods of SEPM for calculating the band structure of Ge/Si QWs and atomistic valence force field for atom position relaxation. From the calculated band structure we obtain the spin splitting and then Rashba parameters of their 2DHGs. Results are shown in Sec.~\ref{section3}. Sec.~\ref{s3-subsection1} shows the electronic structures of Ge/Si QWs; Sec.~\ref{s3-subsection2} shows the $\bf{k}$-linear and $\bf{k}$-cubic Rashba SOC of 2DHGs for QWs with different growth directions; Sec.~\ref{s3-subsection3} and Sec.~\ref{s3-subsection4} show the field- and well width-dependency of the $\bf{k}$-linear and $\bf{k}$-cubic Rashba parameters, respectively. Sec.~\ref{section4} presents the discussion of the results. Sec.~\ref{s4-subsection1} conducts the symmetry analysis of the zone-center HH-LH mixing, revealing the prohibition of HH-LH mixing in even-ML [111]-oriented QWs; Sec.~\ref{s4-subsection2} introduces the interface-inversion-asymmetry induced Dresselhaus SOC in odd-ML [001]-oriented QWs; Sec.~\ref{s4-subsection3} illustrates the local interface induced HH-LH mixing in [001]-oriented and odd-ML [111]-oriented QWs; Sec.~\ref{s4-subsection4} presents the transformation of the LK Hamiltonian and exhibits how the intrinsic HH-LH mixing depends on the QW orientations. Sec.~\ref{section5} summarizes this work.

\section{Computational Methods}\label{section2}

The SEPM calculation uses a supercell approach with a periodic boundary condition, where the supercell contains one Ge/Si QW consisting of $1\times1\times(m+n)$ atomic layers ($m$ and $n$ are Ge and Si thicknesses in the unit of ML). To avoid the interference of the Rashba effect with the BIA-induced Dresselhaus effect, we use Si instead of SiGe alloy to construct the QW barrier to eliminate the Dresselhaus SOC. We adopt the atomistic valence force field (VFF) approach, which has been widely applied to semiconductor nanostructures~\cite{Peter2005, WangLW_PRB2000, Zunger_JAP1998, Luo_PRB2005, Luo_PRB2015, Luo_PRL2012, Luo_NanoLett2012, Luo_NC2013}, to minimize the strain energy in Ge/Si QWs due to a 4.2\% lattice mismatch between bulk Si and Ge (the lattice constants of bulk Si and Ge are 5.43 and 5.65 {\AA}, respectively). All investigated Ge/Si QWs in this work show a 1.3-1.4\% compressive strain in the Ge well and a 2.7-2.8\% tensile strain in the Si barrier. Once obtaining the relaxed atom positions after performing the VFF calculation, we use the atomistic SEPM accompanied with a plane-wave basis set and folded-spectrum diagonalization~\cite{Wang_JCP1994} to obtain the electronic structure of Ge/Si QWs. This set of computational methods has been extensively utilized to study electronic and optical properties of semiconductor superstructures, including  QWs, quantum wires, and quantum dots~\cite{Luo_PRL2012, Luo_NanoLett2012, Luo_PRB2015, Luo2017, Luo_PRL2009, Luo2010, LuoJW_PRL2010, Luo2011, Luo_NatNano2017}. The crystal potential of Ge/Si QWs is a sum of the screened atomic potentials over all Si and Ge atoms within the supercell. The screened Si and Ge atomic potentials include a local part and a non-local spin-orbit interaction part and were obtained by fitting to reproduce experimental transition energies, effective masses, spin-orbit splittings, and deformation potentials of the bulk semiconductors to remove the ``LDA" error in both band gap and effective masses~\cite{Wang1995, Wang1999}. We use an energy cutoff of 8.2 Ry to construct a much smaller plane-wave basis set than first-principles approaches~\cite{Wang1995, Wang1999}.

\section{Results}\label{section3}

\subsection{Electronic structures of Ge/Si QWs}\label{s3-subsection1}

\begin{figure}[!t]
\centering
\includegraphics[width=\linewidth]{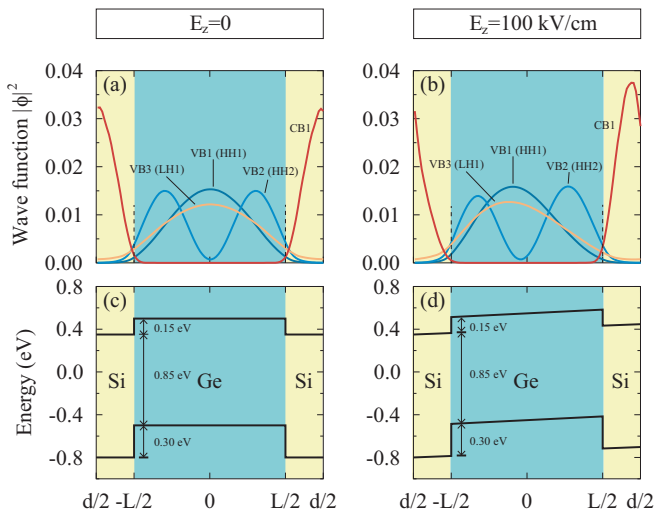}
\caption{(a, b) Normalized wave functions of HH and LH states and (c, d) band offsets of the conduction band minimum (CBM) and the valence band maximum (VBM) in [001]-oriented (Ge)$_{40}$/(Si)$_{20}$ QWs. The external electric field is absent in (a, c) and present in (b, d) with the strength of 100 kV/cm. Here $L$ and $d$ denote the width of the well and the whole QW, respectively. The black dashed line denotes the interface.}
\label{fig1}
\end{figure}

The Ge/Si QWs belong to type-II QWs, where the electrons are confined in Si layers whereas the holes are confined in Ge layers because the valence band maximum (VBM) of bulk Ge is 520 meV higher than that of bulk Si and thus the conduction band minimum (CBM) is also 380 meV higher since the bandgap of the bulk Si and Ge are 1.12 and 0.74 eV, respectively~\cite{Peter2005}. We illustrate in Fig.~\ref{fig1} such type-II band alignment by taking the [001]-oriented (Ge)$_{40}$/(Si)$_{20}$ QW as an example. The VFF-relaxed QW structure has an in-plane lattice constant of 5.58 {\AA}, resulting in a 1.4\% compressive strain in the Ge layers and a 2.7\% tensor strain in the Si layers. Those stains modify the band offsets of the VBM and CBM to 0.30 eV and 0.15 eV, respectively, yielding a QW bandgap of 0.85 eV [Fig.~\ref{fig1}(c, d)]. Fig.~\ref{fig1}(a, b) shows that the band-edge states of the top three valence bands, including HH1, HH2, and LH1, are indeed confined in the Ge layers, whereas that of the first conduction band CB1 is confined in the Si layers. In the absence of the external electric field [i.e., $E_z=0$ shown in Fig.~\ref{fig1}(a, c)], their wave functions are inversion-symmetric and the band alignment is flat. Upon application of an external electric field [say $E_z=100$ kV/cm as shown in Fig.~\ref{fig1}(b, d)], the inversion symmetry of their wave functions is broken with the hole states and electron states shifting towards opposite directions and the band alignment produces a slope. We note that the wave functions at the interface are non-zero [the black dashed line in Fig.~\ref{fig1}(a, b)] due to the finite band offsets [Fig.~\ref{fig1}(c, d)], no matter the external electric field is applied or not. This reflects the importance of the interface effect in QWs, which cannot be ignored in terms of the HH-LH mixing and the Rashba spin splitting, as we will introduce in the rest part of this paper.

\subsection{Linear or cubic Rashba spin splitting in Ge/Si QWs}\label{s3-subsection2}

\begin{figure}[!b]
\centering
\includegraphics[width=\linewidth]{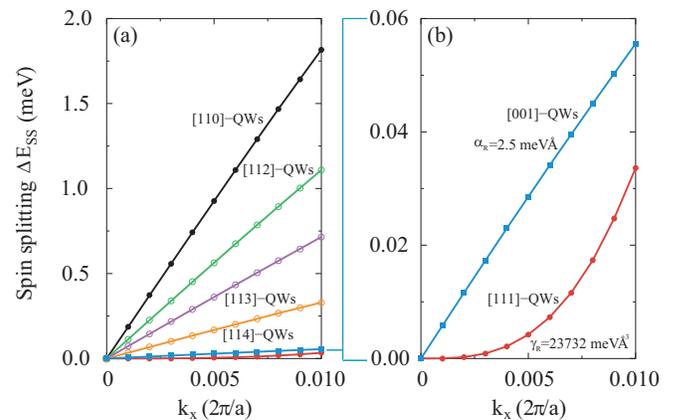}
\caption{(a) Spin splitting of Ge/Si QWs grown in the [110], [111], [112], [113], [114] and [001] crystalline directions with an even number of MLs. (b) Enlarged view on spin splitting of [111]- and [001]-oriented QWs. The $k_x$ directions for different oriented QWs are shown in Table~\ref{tab2}. The external electric field with a strength of 100 kV/cm is applied perpendicularly. The Ge well thicknesses of [110]-, [111]-, [112]-, [113]-, [114]- and [001]-oriented QWs are 40, 48, 40, 48, 32 and 40 monolayers (MLs), respectively, and the Si barrier thicknesses are as half as the Ge well thicknesses.}
\label{fig2}
\end{figure}

We first examine the spin splitting of the first valence band (VB1) of [001]-, [110]-, [111]-, [112]-, [113]-, and [114]-oriented Ge/Si QWs having an even number of MLs in both well and barrier thickness. In the absence of an external electric field, the VB1 shows no spin splitting for all these QWs. Upon application of an external electric field perpendicular to the confinement plane of the QWs, the spin-degeneracy of energy subbands, including VB1, is lifted for wavevector $k$ away from the zone-center $\overline{\Gamma}$ point, giving rise to the SIA-induced Rashba spin splitting. Fig.~\ref{fig2}(a) shows the calculated spin splitting $\Delta E_{ss}$ of the VB1 for [110]-, [111]-, [112]-, [113]-, [114]- and [001]-oriented Ge/Si QWs under an electric field of 100 kV/cm. Here, we use almost the same Ge well thickness of around 80 {\AA} for different growth directions: 40, 48, 40, 48, 32, and 40 MLs for [110]-, [111]-, [112]-, [113]-, [114]-, and [001]-oriented QWs regarding their different interlayer distances. We see that the spin splittings of [110]-, [112]-, [113]-, [114]-, and [001]-oriented QWs exhibit a nice $\bf{k}$-linear dispersion within the sampled $k_x$ range. However, the spin splitting of the [111]-oriented QW possesses a nearly pure $\bf{k}$-cubic dispersion [see Fig.~\ref{fig2}(b)]. Since the diamond structure of both bulk Si and Ge owns an inversion center,  the BIA-induced Dresselhaus effect is expected to be absent, which is indeed confirmed by our atomistic calculations, in these Ge/Si QWs. Thus, the observed spin splitting is exclusively due to the SIA-induced Rashba effect~\cite{Rashba1960}. Such spin splitting $\Delta E_{ss}$ of the VB1 can be formulated as an expansion in powers of the wave vector $\bf{k}$ with only odd terms retained, as required by the time-reversal symmetry. The dominant contributions to $\Delta E_{ss}$ approaching the $\bar{\Gamma}$ point are generally described by the two lowest order terms: $\Delta E_{ss}=2\alpha_Rk+\gamma_Rk^3$, where $\alpha_R$ and $\gamma_R$ indicate the $\bf{k}$-linear and $\bf{k}$-cubic Rashba parameters, respectively. By fitting the spin splitting $\Delta E_{ss}$, we obtain $\alpha_R=81.4,\ 0,\ 49.4,\ 32.9,\ 14.8,\ 2.5$ meV{\AA} for [110]-, [111]-, [112]-, [113]-, [114]- and [001]-oriented QWs, respectively. Since $\alpha_R=0$ in [111]-oriented QWs, we are also interested in the cubic Rashba SOC with obtained parameter $\gamma_R=2.4\times 10^4$ meV\AA$^3$. For [$11n$] crystalline orientations except $n=1$, one can see that the $\bf{k}$-linear Rashba parameter $\alpha_R$ decreases with the increase of Miller index $n$.

\subsection{Electric-field- and well-width-dependence of linear Rashba SOC}\label{s3-subsection3}

\begin{figure}[!t]
\centering
\includegraphics[width=\linewidth]{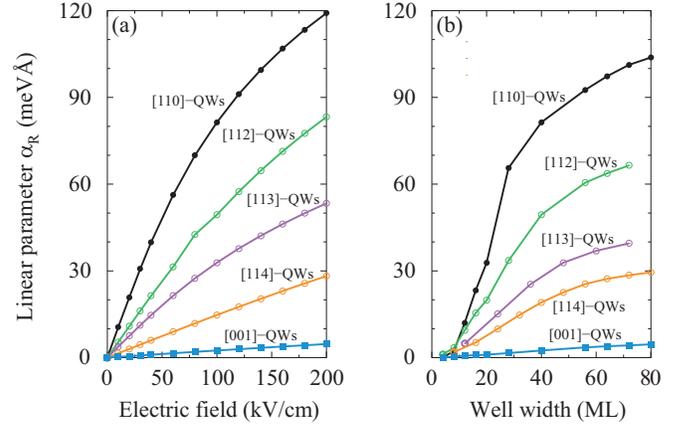}
\caption{Rashba parameters of Ge/Si QWs grown in [110], [112], [113], [114] and [001] crystalline directions as a function of (a) electric field strength with fixed well width (40, 40, 48, 32 and 40 ML), and (b) well width of even MLs under an electric field of 100 kV/cm.}
\label{fig3}
\end{figure}

Fig.~\ref{fig3}(a) shows that the $\bf{k}$-linear Rashba SOC can be effectively tuned by the external electric field for  [110]-, [111]-, [112]-, [113]-, [114]- and [001]-oriented QWs. One can see that, the $\bf{k}$-linear Rashba parameters $\alpha_R$ get larger in a nearly linear relationship for all these QWs as increasing the applied electric field. Particularly, such a linear relationship is perfect in [114]- and [001]-oriented QWs with a relatively weak $\bf{k}$-linear Rashba SOC. As increasing the $\bf{k}$-linear Rashba SOC from [114]-oriented QW to [113]-, [112]- through [110]-oriented QWs, the sub-linear scaling feature becomes more prominent. In the [110]-oriented QW with a strongest  $\bf{k}$-linear Rashba SOC, the sublinear scale is so strong that $\alpha_R$ tends to saturate at a larger electric field. This increased sub-linearity indicates the direct Rashba effect and reflects the increasing quantum confinement Stark effect~\cite{Luo2017, Xiong_PRB2021}, which is induced by the external electric field with the same strength.

Fixing the applied external electric field to 100 kV/cm, we further investigate the strength of the Rashba SOC by varying the Ge well thickness for investigated QWs. Fig.~\ref{fig3}(b) exhibits that the $\bf{k}$-linear Rashba parameters $\alpha_R$ grow as the well becomes wider for  [110]-, [111]-, [112]-, [113]-, [114]- and [001]-oriented QWs. Such well-width dependence is a fingerprint of the direct Rashba SOC~\cite{Luo2011, Xiong_PRB2021, Kloeffel2011}. According to the conventional Rashba SOC, the $\bf{k}$-linear term is vanishing and the $\bf{k}$-cubic term was believed to be the lowest order in semiconductor 2DHGs~\cite{Winkler2000, Winkler2003, Bulaev2005, EDSR2007, Moriya2014, Marcellina_PRB2017}. Strictly speaking, there is also a $\bf{k}$-linear term in the conventional Rashba SOC stemming from the $k\cdot p$ coupling between the bonding and anti-bonding atomic $p$ orbitals (the latter gives rise to the first excited conduction band)~\cite{Marcellina_PRB2017}. This $\bf{k}$-linear term is usually neglected for its less than 1\% contribution to the total spin splitting. Additionally, the bulk $\bf{k}$-cubic term of the conventional Rashba SOC will also give rise to a $\bf{k}$-linear Rashba term in QWs as a result of the quantization of wave vector in the QW growth direction: $\alpha_R\sim \gamma_R<\hat{k}_z^2>$. Because of the substitution of the wave vector operator $\hat{k}_z\sim\pi/L$, the linear Rashba parameter $\alpha_R$ should be inversely proportional to the well width, which is opposite to the observed well-width-dependence of  $\alpha_R$ [shown in Fig.~\ref{fig3}(b)]. Indeed, such $\bf{k}$-linear Rashba SOC is also negligibly small~\cite{Kloeffel2011}. Different from the conventional Rashba effect, here, the emerging $\bf{k}$-linear spin splitting should originate from a direct dipolar coupling of HH subbands to the external electric field in the presence of HH-LH mixing at zone-center~\cite{Xiong_PRB2021} and is thus called the direct Rashba effect~\cite{Kloeffel2011, Kloeffel2018a, Kloeffel_PRB2013, Luo2017, Xiong_PRB2021}. In terms of the direct Rashba SOC theory,  for [001]-oriented QWs we obtain a $\bf{k}$-linear Rashba parameter $\alpha_R$ as following~\cite{Xiong_PRB2021}
\begin{equation}\label{Rashba2}
\alpha_R^{[001]}=\frac{2e\gamma_3C_0U_0E_z}{\sqrt{\Delta_0^2+4e^2U_0^2E_z^2}}.
\end{equation}
It is due to the external electric field $E_z$ that couples directly to the spins if the HH-LH mixing is allowed in QWs, yielding a direct dipolar coupling term $\langle \textrm{HH1}_{\pm}|(-eE_zz)|\textrm{HH2}_{\pm} \rangle = eE_zU_0$, where the coupling constant $U_0=(a_1a_2+b_1b_2)16L/(9\pi^2)$ is also related to the HH-LH mixing~\cite{Xiong_PRB2021}. It is straightforward to learn that $\bf{k}$-linear Rashba parameter $\alpha_R$ of the direct Rashba SOC is larger for a wider well.

\subsection{Electric-field- and well-width-dependence of cubic Rashba SOC}\label{s3-subsection4}

\begin{figure}[!t]
\centering
\includegraphics[width=\linewidth]{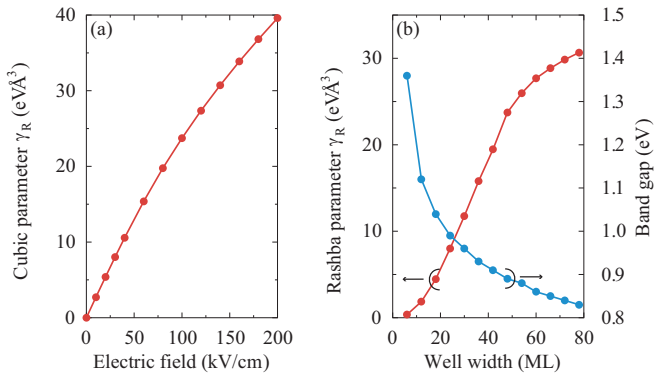}
\caption{Rashba parameters of Ge/Si QWs grown in [111] crystalline direction as a function of (a) electric field strength with fixed well width (48 ML), and (b) well width of even MLs under an electric field of 100 kV/cm.}
\label{fig4}
\end{figure}

In [111]-oriented QWs, the $k$-linear Rashba SOC is fully absent with a purely k-cubic spin splitting in energy subbands. It indicates the absence of direct Rashba effect in [111]-oriented QWs. Thus, this $\bf{k}$-cubic Rashba SOC is a conventional Rashba effect produced by the second-order perturbation and has an inverse proportion to the square of the bandgap. Fig.~\ref{fig4}(a) shows that the $\bf{k}$-cubic Rashba parameter $\gamma_R$ in even-ML [111]-oriented QWs has a linear scale as applied electric field $E_z$, satisfying the theoretical prediction of linear-in-field behavior in $\bf{k}$-cubic Rashba parameters~\cite{Winkler2003}. The $\bf{k}$-cubic Rashba parameter $\gamma_R$ can be enhanced to 40 eV{\AA}$^3$ at E$_z$=200 kV/cm. Under a fixed electric field E$_z$=100 kV/cm, Fig.~\ref{fig4}(b) shows that $\gamma_R$ increases linearly against the well width when $L<48$ ML, and then tends to saturate to a steady value as further increasing the well width. It is well known that when the well width becomes larger, the quantum confinement effect will be reduced, making the bandgap decrease and gradually saturate, inconsistent with the blue line shown in Fig.~\ref{fig4}(b). Because the $\bf{k}$-cubic Rashba parameters $\gamma_R$ is reversely proportional to the square of the band gap ($\propto 1/E_g^2$)~\cite{Winkler2003}, $\gamma_R$ will increase till saturation.

\section{Discussion}\label{section4}

We have demonstrated that the $\bf{k}$-linear Rashba SOC of 2DHGs in [110]-, [111]-, [112]-, [113]-, [114]- and [001]-oriented QWs is a direct Rashba SOC with its Rashba parameter $\alpha_R$ being proportional to the strength of zone-center HH-LH mixing. The strongest Rashba SOC (largest $\alpha_R$) occurred in the [110]-oriented QWs among investigated growth orientations indicates that the [110]-oriented QWs possess the largest zone-center HH-LH mixing. Meanwhile, the absence of linear spin splitting in the [111]-oriented QWs indicates that the HH-LH mixing is forbidden therein. To understand the orientation-dependent linear or cubic Rashba SOC, one should figure out two essential questions on HH-LH mixing: (i) How does the symmetry govern the presence or absence of the zone-center HH-LH mixing in QWs? (ii) If HH-LH mixing is allowed, what is the factor regulating their strength in QWs? We will address the first question in Sec.~\ref{s4-subsection1} and Sec~\ref{s4-subsection2}, and the second question in Sec.~\ref{s4-subsection3} and Sec.~\ref{s4-subsection4}, respectively.

\subsection{Symmetry analysis of zone-center HH-LH mixing}\label{s4-subsection1}

\begin{table*}[!t]
\caption{Double point groups of the Ge/Si QWs in combination of the counterparts of the local interface with corresponding irreducible representations of HH and LH states for different growth directions~\cite{Altmann_PG, Winkler2003, Satpathy_PRB1988, Ivchenko1996, Golub2004, Cartoixa2006}. Here we adopt the Koster notations for the irreducible representations of point groups.}
\begin{tabular}{p{1.0 cm}<{\centering} | p{2.0 cm}<{\centering} | p{2.0 cm}<{\centering} | p{2.0 cm}<{\centering} | p{2.0 cm}<{\centering} | p{2.0 cm}<{\centering} | p{2.0 cm}<{\centering}}
  \hline\hline
  \multicolumn{1}{c|}{\multirow{2}{*}{QW orientations}} & \multicolumn{2}{c|}{[001]} & \multicolumn{2}{c|}{[110]} & \multicolumn{2}{c}{[111]} \\ \cline{2-7}
   & odd & even & odd & even & odd & even  \\
  \hline
  \multicolumn{1}{c|}{Global point groups of QWs} & $D_{2d}$ & $D_{2h}$ & $D_{2h}$ & $D_{2h}$ & $C_{3v}$ & $D_{3d}$ \\ \hline
  \multicolumn{1}{c|}{HH} & $\Gamma_6$ & $\Gamma_5$ & $\Gamma_5$ & $\Gamma_5$ & $\Gamma_6$ & $\Gamma_4$, $\Gamma_5$ \\ \cline{1-7}
  \multicolumn{1}{c|}{LH} & $\Gamma_7$ & $\Gamma_5$ & $\Gamma_5$ & $\Gamma_5$ & $\Gamma_6$ & $\Gamma_8$ \\
  \hline
  \multicolumn{1}{c|}{Point groups of the local interface} & $C_{2v}$ & $C_{2v}$ & $C_{2v}$ & $C_{2v}$ & $C_{3v}$ & $C_{3v}$ \\ \hline
  \multicolumn{1}{c|}{HH} & $\Gamma_5$ & $\Gamma_5$ & $\Gamma_5$ & $\Gamma_5$ & $\Gamma_6$ & $\Gamma_4$, $\Gamma_5$ \\ \cline{1-7}
  \multicolumn{1}{c|}{LH} & $\Gamma_5$ & $\Gamma_5$ & $\Gamma_5$ & $\Gamma_5$ & $\Gamma_6$ & $\Gamma_6$ \\
  \hline\hline
\end{tabular}
\label{tab1}
\end{table*}

Bulk Si and Ge possess an $O_h$ point group, in which the HH and LH bands at the $\Gamma$ point belong to the four-dimensional $\Gamma_8^+$ representation~\cite{Altmann_PG} and thus are degenerate (including spin). In Ge/Si QWs, the crystal system reduces from bulk $O_h$ to a lower-symmetric point group in which the zone-center HH and LH states will certainly transform according to different irreducible representations from bulk $\Gamma_8^+$. The confinement potential of the QWs deviating from bulk crystals may mix bulk HH and LH states if and only if they belong to the same representation. This is because the perturbation crystal potential from the bulk crystals cannot couple two states with different irreducible representations due to the symmetry of the crystal potential in the ground state belonging to the $\Gamma_1$ representation. In the [001]-oriented GaAs/AlAs QWs with a $D_{2d}$ point group, the irreducible representations of the zone-center HH and LH states reduce from $\Gamma_8$ in bulk Zinc-blende $T_d$ point group to $\Gamma_6$ and $\Gamma_7$ of the $D_{2d}$ point group, respectively. Thus, the HH-LH mixing is forbidden by symmetry. However, there is both experimental and theoretical evidence showing the finite zone-center HH-LH mixing in the [001]-oriented GaAs/AlAs QWs. It was argued that such HH-LH mixing is induced by the $C_{2v}$ local interface although it was forbidden by the $D_{2d}$ global symmetry of the QW~\cite{Winkler2003, Golub2004, Ivchenko1996}.  The local interface in the [001]-oriented GaAs/AlAs QWs has a $C_{2v}$ symmetry due to the lack of microscopic translational symmetry since the atoms at the two sides of the interface belong to different elements~\cite{Winkler2003, Golub2004, Ivchenko1996}. In the $C_{2v}$ point group, the HH and LH belong to the same $\Gamma_5$ irreducible representation and thus are allowed to couple. We therefore have to examine both the global point groups of the QWs and point groups of their local interface. Table~\ref{tab1} summaries the corresponding irreducible representations of HH and LH states in [001]-, [110]-, and [111]-oriented Ge/Si QWs~\cite{Altmann_PG, Winkler2003, Satpathy_PRB1988, Ivchenko1996, Golub2004, Cartoixa2006}.

\begin{figure}[!b]
\centering
\includegraphics[width=\linewidth]{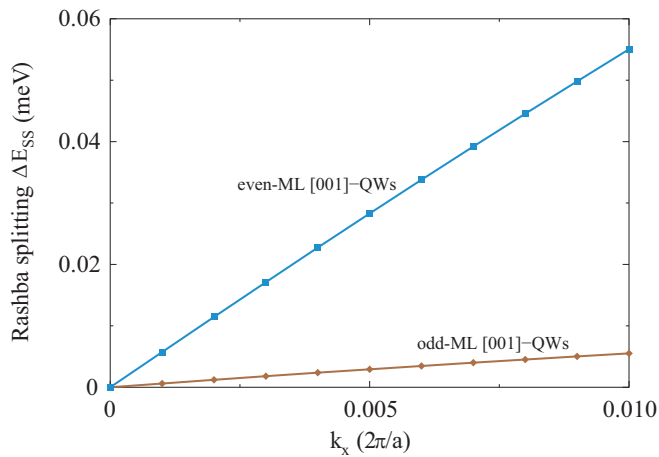}
\caption{(a) Rashba spin splitting of [001]-oriented Ge/Si QWs with an odd and even number of the well MLs. The external electric field with a strength of 100 kV/cm is applied perpendicularly. The Ge well thicknesses of the odd and even number of MLs are 39 and 40, and the Si barrier thicknesses are 21 and 20 MLs, repectively.}
\label{fig5}
\end{figure}

Because of the point-group symmetry differences between an odd and even number of MLs in Ge$_n$/Si$_m$, besides the even-ML QWs discussed above, we also include the symmetry of odd-ML QWs in Table~\ref{tab1}.  One can see from Table~\ref{tab1} that the global symmetry of the [001]-oriented Ge/Si QWs is a $D_{2h}$ point group when both Ge and Si layers have an even number of MLs, or a $D_{2d}$ point group when both Ge and Si layers have an odd number of MLs~\cite{Golub2004, Nestoklon2008}. It is in sharp contrast to [001]-oriented GaAs/AlAs QWs where the QW point group is always $D_{2d}$ as varying the layer thickness~\cite{Kitaev1997}. The local symmetry of interfaces in both even- and odd-ML [001]-oriented QWs is a $C_{2v}$ point group. Therefore, the odd-ML [001]-oriented Ge/Si QWs have the same feature of zone-center HH-LH mixing as in both even-ML and odd-ML group III-V QWs: the zone-center HH-LH mixing is forbidden by the QW global $D_{2d}$ point group but allowed by the local interface with $C_{2v}$ symmetry under which both HH and LH states transform according to the same $\Gamma_5$ representation and thus the HH-LH mixing is allowed by symmetry~\cite{Ivchenko1996}. However, this situation is changed in even-ML Ge/Si QWs. Table~\ref{tab1} shows that, in both $D_{2h}$  and $C_{2v}$ point groups, both the zone-center HH and LH states transform according to the $\Gamma_5$ representation. Therefore, in even-ML [001]-oriented Ge/Si QWs the zone-center HH-LH mixing is allowed by QW global symmetry and interface local $C_{2v}$ symmetry in which both QW confinement potential and local interface potential could induce HH-LH mixing. Fig.~\ref{fig5} shows that both odd-ML and even-ML [001]-oriented Ge/Si QWs have linear Rashba spin splitting but the odd-ML QW is much smaller than the even-ML one, which implies that the QW global symmetry allowed HH-LH mixing remarkably enhances the $\bf{k}$-linear Rashba SOC in the even-ML QW. We should note that two interfaces in the odd-ML Ge/Si QWs break the inversion symmetry, introducing the interface-inversion-asymmetry (IIA) induced Dresselhaus spin splitting~\cite{Nestoklon2008} which has been subtracted from that shown in Fig.~\ref{fig5} to obtain the Rashba spin splitting. We will discuss it in Sec.~\ref{s4-subsection2} in detail.

In both even-ML and odd-ML [110]-oriented Ge/Si QWs, the point group of the QW global crystal is $D_{2h}$ and the interface has a local $C_{2v}$ point group, as given in Table~\ref{tab1}. Same as in even-ML [001]-oriented Ge/Si QWs, both the QW global $D_{2h}$ symmetry and interface local $C_{2v}$ symmetry allow the zone-center HH-LH mixing since both HH and LH states transform according to the same $\Gamma_5$. Therefore, the QW confinement potential and interface local potential could induce HH-LH mixing in [110]-oriented Ge/Si QWs. Our atomistic calculations also show that [110]-oriented (Ge)$_{40}$/Si$_{20}$ (even-ML) and (Ge)$_{39}$/(Si)$_{21}$ (odd-ML) QWs under an electric field of 100 kV/cm have very similar linear Rashba spin splitting with linear Rashba parameter of $\alpha_R=81.4$  and 81.1 meV{\AA}, respectively.

\begin{figure}[!t]
\centering
\includegraphics[width=\linewidth]{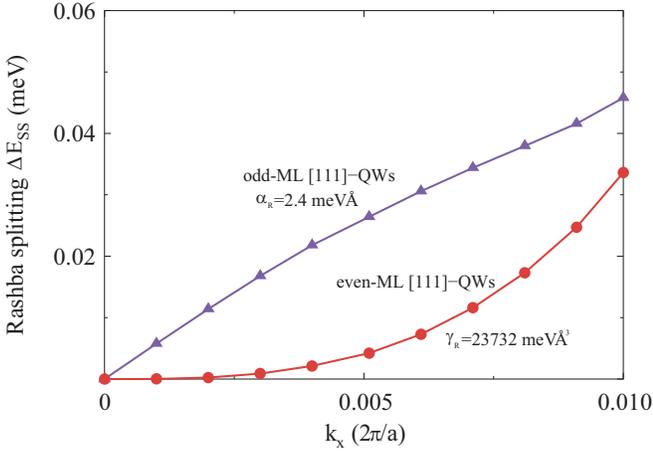}
\caption{(a) Rashba spin splitting of [111]-oriented Ge/Si QWs with an odd and even number of the well MLs. The external electric field with a strength of 100 kV/cm is applied perpendicularly. The Ge well thicknesses of the odd and even number of MLs are 47 and 48, and the Si barrier thicknesses are 25 and 24 MLs, repectively.}
\label{fig6}
\end{figure}

In the [111]-oriented QWs, the QW global symmetry is $D_{3d}$ for even-ML case and is $C_{3v}$ for odd-ML case. The local symmetry of interface is $C_{3v}$ in both cases.  In the even-ML [111]-oriented QWs, the two spin components of the zone-center HH state transform according to $\Gamma_4$ and $\Gamma_5$ representations, respectively, and the LH state belongs to the $\Gamma_8$ representation in QW global $D_{3d}$ point group. Meanwhile, in the interface local $C_{3v}$ point group, the HH states belong to $\Gamma_4$ and $\Gamma_5$ representations (for two spin components), and the LH states belong to the $\Gamma_6$ representation~\cite{Cartoixa2006}. Therefore, in even-ML [111]-oriented Ge/Si QWs, the HH-LH mixing is forbidden both by QW global symmetry and interface local symmetry, leading to the absence of direct Rashba SOC and leaving alone the conventional Rashba SOC. It thus explains why we have purely $k$-cubic Rashba spin splitting in the even-ML [111]-oriented QWs, as shown in Fig.~\ref{fig2}(b). Interestingly, in the odd-ML [111]-oriented Ge/Si QWs, the representations of both zone-center HH and LH states reduce to the same $\Gamma_6$ in $C_{3v}$ point group and thus the HH-LH mixing is allowed by both QW global symmetry and interface local symmetry. Hence, both QW confinement potential and interface local potential might induce HH-LH mixing and thus $\bf{k}$-linear direct Rashba SOC. Fig.~\ref{fig6} exhibits that the odd-ML [111]-oriented Ge/Si QWs has a linear spin splitting in striking contrast to the purely $k$-cubic spin splitting in the even-ML case.

\subsection{Interface-inversion-asymmetry induced Dresselhaus SOC}\label{s4-subsection2}

\begin{figure}[!t]
\centering
\includegraphics[width=\linewidth]{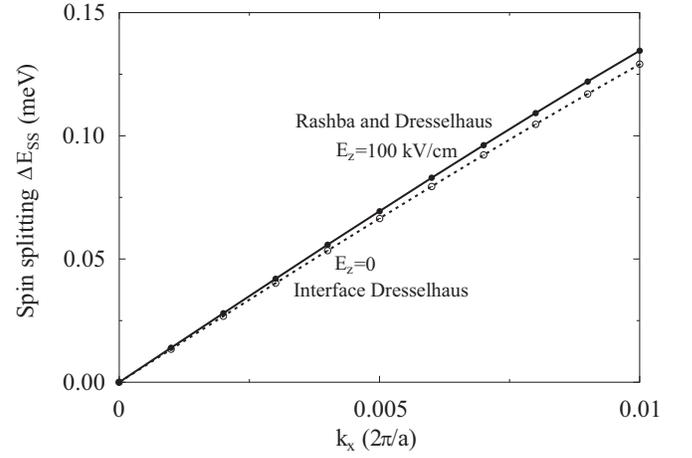}
\caption{Spin splitting of odd-ML [001]-oriented (Ge)$_{39}$/(Si)$_{21}$ QWs along $k_x$ direction with (the black solid line) or without an external electric field of 100 kV/cm (the black dashed line).}
\label{fig7}
\end{figure}

Although both bulk Si and Ge are in diamond crystal structure with an inversion center, the inversion symmetry is broken due to the interfaces in the odd-ML [001]-oriented Ge/Si QWs, which possess a non-centrosymmetric point group $D_{2d}$. Hence, as shown in Fig.~\ref{fig7}, the [001]-oriented (Ge)$_{39}$/(Si)$_{21}$ QW has a spin splitting even without external electric field. Because the inversion asymmetry in the absence of external electric field arises from the atom arrangement on the interfaces, the zero-field spin splitting is termed as the IIA induced Dresselhaus effect~\cite{Golub2004, Nestoklon2008, Winkler2003}. The zero-field Dresselhaus spin splitting exhibits a linear function against the wave vector $k$ with the linear Dresselhaus parameter $\beta_D$ fitted to be 6.0 meV{\AA}.

Upon application of an external electric field, the electric-field-induced direct Rashba effect will also contribute to the spin splitting.  Fig.~\ref{fig7} shows that a 100 kV/cm electric field enhances the spin splitting and increases the $k$-linear parameter to 6.2 meV{\AA}. The difference in the spin splitting of the two above cases with and without external electric field is the $k$-linear direct Rashba SOC (the brown line in Fig.~\ref{fig5}). This $\bf{k}$-linear Rashba SOC in odd-ML [001]-oriented QWs results completely from the interface local $C_{2v}$ symmetry induced HH-LH mixing, which is nominally forbidden by the QW global $D_{2d}$ symmetry.

\subsection{Local interface induced HH-LH mixing}\label{s4-subsection3}

Having illuminated the presence or absence of the zone-center HH-LH mixing being governed by the local interface symmetry rather than QW global symmetry, we next proceed to address the second question of which factor regulates mainly the strength of HH-LH mixing in QWs with respect to growth directions. According to the symmetry analysis, we can separate the contributions to the HH-LH mixing into two types of sources. The first type source arises from the breaking of the axial symmetry~\cite{Winkler2003} and is irrelevant to both the local interface and external electric field, depending only on the QW growth direction. This type of HH-LH mixing is allowed by the global symmetry of QWs and thus usually exists in low symmetry QWs~\cite{Winkler2003}. Here we call this type of HH-LH mixing as intrinsic HH-LH mixing. The second type of HH-LH mixing is induced by the local interface, as suggested firstly by Ivchenko et al.~\cite{Ivchenko1996} to explain the observed HH-LH mixing in group III-V semiconductor [001]-oriented QWs in which the HH-LH mixing is forbidden by QW global symmetry. The interface-induced HH-LH mixing has no direct connection to the QW confinement potential, and hence is expected to be weak relative to the intrinsic HH-LH mixing (if allowed by symmetry) because the wave functions of band-edge states localize inside the QW with only a tiny component on the interface as illustrated in Fig.~\ref{fig1}(a) and Fig.~\ref{fig1}(b). Thus, the $\bf{k}$-linear Rashba SOC in the [001]-oriented QW (Fig.~\ref{fig5}) and the odd-ML [111]-oriented QW (the purple line in Fig.~\ref{fig6}) is completely correlated with the interface-induced HH-LH mixing, whereas that of [110]-oriented QW arises both from the intrinsic and interface-induced HH-LH mixing. There, the fact that the linear Rashba parameter $\alpha_R$ is one order of magnitude larger in [110]-oriented QWs than that in [001]-oriented QWs might confirm that the interface-induced HH-LH mixing is negligible~\cite{Winkler2003} compared to the intrinsic HH-LH mixing.

It is worthy to note that the argument that the interface local symmetry alone could induce HH-LH mixing even if the QW global symmetry forbids it remains lacking direct theoretical proof~\cite{Winkler2003, Golub2004, Ivchenko1996}. Here, the absence of the $\bf{k}$-linear Rashba SOC in even-ML [111]-oriented QWs (the red line in Fig.~\ref{fig2} and Fig.~\ref{fig6}) in combination with the finite $\bf{k}$-linear Rashba SOC occurred in odd-ML [001]-oriented QWs (the brown line in Fig.~\ref{fig5}) provides direct proof for this hypothesis. Specifically, we have demonstrated that the linear direct Rashba SOC is proportional to the strength of the HH-LH mixing and that the purely cubic Rashba spin splitting in the even-ML [111]-oriented QWs is due to the absence of the direct Rashba SOC. We thus learn that the HH-LH mixing occurs in the odd-ML [001]-oriented QWs but lacks in the even-ML [111]-oriented QWs. In the former case, the HH-LH mixing is forbidden by the QW $D_{2d}$ global symmetry but allowed by the local $C_{2v}$ interface symmetry. Whereas, in the latter case both the QW $D_{3d}$ global symmetry and the local $C_{3v}$ interface symmetry forbid the HH-LH mixing. Consequently, only the local interface symmetry can tell the absence and presence of HH-LH mixing (hence linear Rashba spin splitting) in even-ML [111]-oriented QWs and odd-ML [001]-oriented QWs, respectively. Besides the linear Rashba SOC, the local interface symmetry in QWs can also be utilized to address other physics related to the HH-LH mixing, such as effective mass~\cite{Cheng2021}, resonant tunneling~\cite{Schuberth1991}, and optical excitation~\cite{Belhadj2021, Huebers2005}.

\subsection{Assess the QW confinement potential induced intrinsic HH-LH mixing}\label{s4-subsection4}

\begin{figure*}[!t]
\centering
\includegraphics[width=\linewidth]{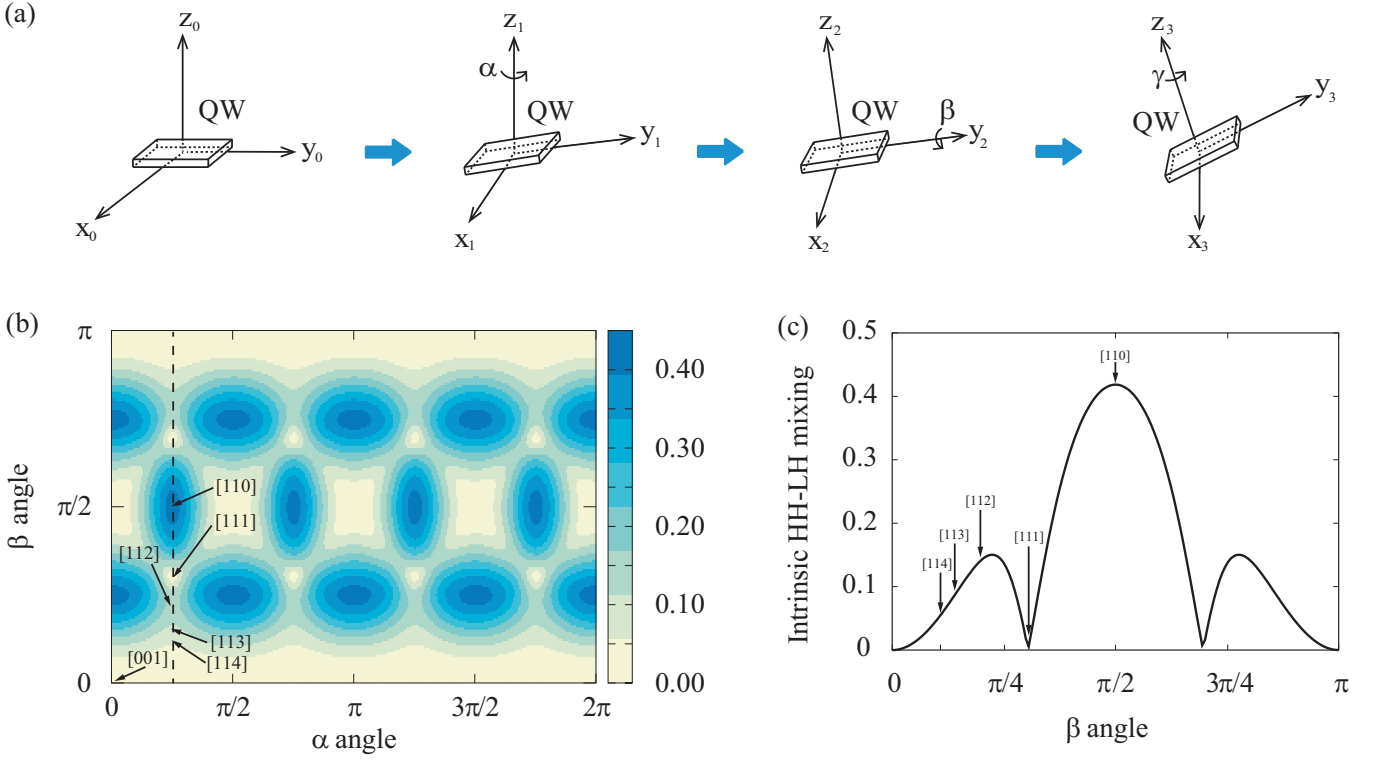}
\caption{(a) The schematic diagram of the coordinate system transformation of a QW, including three steps: rotation around the z-axis with $\alpha$ angle, rotation around the y-axis with $\beta$ angle, and rotation around the z-axis with $\gamma$ angle. (b) The intrinsic HH-LH mixing with different QW orientations, where the z-direction is determined by the Euler angles $\alpha$ and $\beta$. (c) The intrinsic HH-LH mixing with $\alpha=\frac{\pi}{4}$ as a function of the $\beta$ angle, i.e., along with the [$mmn$] crystalline direction, which is marked as a dashed line in (b).}
\label{fig8}
\end{figure*}

From symmetry analysis we learn that in [110]-oriented QWs and even-ML [001]-oriented QWs, both QW global symmetry and local interface symmetry allow the HH-LH mixing, but the former has the strongest linear Rashba SOC with one order of magnitude larger in linear Rashba parameter $\alpha_R$ than the latter. It manifests that the confinement potential induced intrinsic HH-LH mixing must be very sensitive to the varying QW growth direction. The intrinsic HH-LH mixing, if present, can be assessed from the off-diagonal matrix terms in the LK Hamiltonian, which describes valence bands of semiconductors with hole spins taken into consideration. Because the LK Hamiltonian describes the bulk system, we should quantize one direction (say the $z$ direction) to represent the quantum confinement effect. The quantization of the $z$ direction is accompanied by the symmetry reduction from bulk to QW systems. The symmetry of bulk crystal ($O_h$ group) contains all the axial, cubic, and tetrahedral rotation symmetry, whereas the QW symmetry (continuous group $D_{\infty h}$)  possesses only the axial rotation symmetry~\cite{Winkler2003}. For QWs with different orientations, we transform the bulk LK Hamiltonian to make sure its $z$ direction is along the growth direction of the QW and quantize the wave vector along the $z$ direction (i.e., the growth direction) to obtain the QW Hamiltonian. One should note that this transformation of the LK Hamiltonian associated with the quantization of the $z$ direction remains uninvolved with any information from the interface effect, which can be described by the interface Hamiltonian $H_{int}$ proportional to $\delta(z-z_i)$ ($z_i$ denotes the location of the interface)~\cite{Ivchenko1996, Winkler2003}. The form of the interface Hamiltonian $H_{int}$, associated with the envelope functions, depends on the boundary conditions~\cite{Durnev_PRB2014, Marcellina_PRB2017, Ivchenko1996}. Once the boundary condition is given, the symmetry of the total Hamiltonian $H_{tot}=H_{LK}+H_{int}$ will reduce from $D_{\infty h}$ to concrete point groups $D_{2d}$, $D_{2h}$ and $D_{3d}$~\cite{Winkler2003}, as shown in Table~\ref{tab1}. Hence, the total Hamiltonian satisfies the symmetry requirements of different QW orientations.

In the cartesian coordinate system with $x^\prime\parallel$[100], $y^\prime\parallel$[010], $z^\prime\parallel$[001], the bulk LK Hamiltonian~\cite{Luttinger1956} is written as
\begin{widetext}
\begin{equation}\label{LK Hamiltonian}
\begin{split}
H_{LK}=&\frac{\hbar^2}{2m_0}[(\gamma_1+\frac{5}{2}\gamma_2)(k_{x^\prime}^2+k_{y^\prime}^2+k_{z^\prime}^2)-2\gamma_2(k_{x^\prime}^2J_{x^\prime}^2+k_{y^\prime}^2J_{y^\prime}^2+k_{z^\prime}^2J_{z^`}^2) \\
&-4\gamma_3(\{k_{x^\prime},k_{y^\prime}\}\{J_{x^\prime},J_{y^\prime}\}+\{k_{y^\prime},k_{z^\prime}\}\{J_{y^\prime},J_{z^\prime}\}+\{k_{z^\prime},k_{x^\prime}\}\{J_{z^\prime},J_{x^\prime}\})],
\end{split}
\end{equation}
where $\gamma_1$, $\gamma_2$ and $\gamma_3$ are the Luttinger parameters, $\hbar$ the reduced Plank constant, $m_0$ the bare electron mass, $k_i\ (i=x^\prime, y^\prime, z^\prime)$ the wave vectors, $J_i\ (i=x^\prime, y^\prime, z^\prime)$ the hole effective spin operators, and $\{\hat{A},\hat{B}\}=\frac{\hat{A}\hat{B}+\hat{B}\hat{A}}{2}$. To transform the coordinate system, we use the rotation operator $\hat{R}(\alpha,\beta,\gamma)$ defined as
\begin{equation}\label{rotation}
\hat{R}(\alpha,\beta,\gamma)=\left (
\begin{array}{ccc}
\cos\alpha \cos\beta \cos\gamma-\sin\alpha \sin\gamma & -\cos\alpha \cos\beta \sin\gamma-\sin\alpha \cos\gamma & \cos\alpha \sin\beta \\
\sin\alpha \cos\beta \cos\gamma+\cos\alpha \sin\gamma & -\sin\alpha \cos\beta \sin\gamma+\cos\alpha \cos\gamma & \sin\alpha \sin\beta \\
-\sin\beta \cos\gamma & \sin\beta \sin\gamma & \cos\beta \\
\end{array}
\right ),
\end{equation}
where $\alpha$ ($0\le\alpha\le 2\pi$), $\beta$ ($0\le\beta\le\pi$) and $\gamma$ ($0\le\gamma\le 2\pi$) are the Euler angles denoting the rotation angles in order of the $z^\prime$, $y^\prime$ and $z^\prime$ axes resplectively, as shown in Fig.~\ref{fig8}(a). We need three steps to transform the initial coordinate system into the final coordinate system: (i). rotation around the $z$-axis with $\alpha$ angle ($0\le\alpha\le 2\pi$); (ii). rotation around the $y$-axis with $\beta$ angle ($0\le\beta\le\pi$); (iii). rotation around the $z$-axis with $\gamma$ angle ($0\le\gamma\le 2\pi$). The relation between these two basis vectors can be written as

\begin{equation}\label{transformed basis}
\begin{split}
   \left(
   \begin{array}{ccc}
     \hat{e}_x, \hat{e}_y, \hat{e}_z \\
   \end{array}
   \right)=\hat{R}(\alpha,\beta,\gamma)
   \left(
   \begin{array}{ccc}
     \hat{e}_{x^\prime}, \hat{e}_{y^\prime}, \hat{e}_{z^\prime} \\
   \end{array}
   \right).
\end{split}
\end{equation}

By doing this transformation, we can obtain the transformed LK Hamiltonian in the form of the Euler angles $\alpha$, $\beta$ and $\gamma$. To obtain the QW Hamiltonian, we quantize the $z$ direction (the confinement direction) of this Hamiltonian by setting $k_z=\frac{\pi}{L}$, assuming an infinitely deep rectangular well, where $L$ denotes the well width. The transformed LK Hamiltonian is quite complicated, but we can simplify it according to our interests. Because we are only interested in the intrinsic HH-LH mixing at $\bar{\Gamma}$ point, we set $k_x=k_y=0$ in this Hamiltonian. The states $|\frac{3}{2}, \pm\frac{3}{2}\rangle$ and $|\frac{3}{2}, \pm\frac{1}{2}\rangle$ represent the HH and LH states respectively, where the plus and minus signs are two different ``spin blocks". In the basis of {$|J^2, J_z\rangle=\{|\frac{3}{2}, \frac{3}{2}\rangle, |\frac{3}{2}, -\frac{3}{2}\rangle, |\frac{3}{2}, \frac{1}{2}\rangle, |\frac{3}{2}, -\frac{1}{2}\rangle$\}}, the Hamiltonian matrix elements are

\begin{equation}\label{Hamiltonian matrices}
\begin{split}
(H_{LK})_{11}=&\frac{\pi^2\hbar^2[32\gamma_1-31\gamma_2-33\gamma_3-3(\gamma_2-\gamma_3)(4\cos2\beta+7\cos4\beta+8\cos4\alpha\sin^4\beta)]}{64m^*L^2} \\
(H_{LK})_{12}=&0 \\
(H_{LK})_{13}=&\frac{\sqrt{3}\pi^2\hbar^2(\gamma_2-\gamma_3)e^{i\gamma}\sin\beta[(9-\cos4\alpha)\cos\beta+(7+\cos4\alpha)\cos3\beta-4i\sin4\alpha\sin^2\beta]}{16m^*L^2} \\
(H_{LK})_{14}=&-\frac{\sqrt{3}\pi^2\hbar^2(\gamma_2-\gamma_3)e^{2i\gamma}\sin^2\beta[5+7\cos2\beta+\cos4\alpha(3+\cos2\beta)+4i\cos\beta\sin4\alpha]}{16m^*L^2},
\end{split}
\end{equation}
Because only the HH-LH mixing of two different spin blocks ($|\frac{3}{2}, \frac{3}{2}\rangle$ and $|\frac{3}{2}, -\frac{1}{2}\rangle$) contributes to the $\bf{k}$-linear Rashba SOC~\cite{Kloeffel2011, Kloeffel_PRB2013, Kloeffel2018a, Xiong_PRB2021}, the matrix element $H_{14}$ is proportional to the HH-LH mixing. Considering other couplings related to the state $|\frac{3}{2}, \frac{3}{2}\rangle$ and ruling out the influence of the well width, we renormalize the intrinsic HH-LH mixing in the form of
\begin{equation}\label{mixing}
C^{[mnl]}_{in}=\frac{|(H_{LK})_{14}|}{\sqrt{(H_{LK})_{11}^2+(H_{LK})_{12}^2+(H_{LK})_{13}^2+(H_{LK})_{14}^2}}.
\end{equation}

\begin{table}[!t]
\caption{Euler angles and directions of the wave vector basis for different oriented QWs.}
\begin{tabular}{p{2.4 cm}<{\centering} | p{2.4 cm}<{\centering} | p{2.4 cm}<{\centering} | p{2.4 cm}<{\centering} | p{2.4 cm}<{\centering} | p{2.4 cm}<{\centering} | p{2.4 cm}<{\centering}}
  \hline\hline
  QW Orientation & [110] & [111] & [112] & [113] & [114] & [001] \\ \hline
  $k_x$ direction & [001] & [1$\bar{1}$0] & [1$\bar{1}$0] & [12$\bar{1}$] & [1$\bar{1}$0] & [100] \\ \hline
  $k_y$ direction & [1$\bar{1}$0] & [11$\bar{2}$] & [11$\bar{1}$] & [$\bar{7}$41] & [22$\bar{1}$] & [010] \\ \hline
  $\alpha$ & $\frac{\pi}{4}$ & $\frac{\pi}{4}$ & $\frac{\pi}{4}$ & $\frac{\pi}{4}$ & $\frac{\pi}{4}$ & 0 \\ \hline
  $\beta$ & $\frac{\pi}{2}$ & $arccos(\frac{\sqrt{3}}{3})$ & $arccos(\frac{\sqrt{6}}{3})$ & $arccos(\frac{3\sqrt{11}}{11})$ & $arccos(\frac{2\sqrt{2}}{3})$ & 0 \\ \hline
  $\gamma$ & $\pi$ & $\frac{3}{2}\pi$ & $\frac{3}{2}\pi$ & $arctan(\frac{\sqrt{11}}{11})$ & $\frac{3}{2}\pi$ & 0 \\
  \hline\hline
\end{tabular}
\label{tab2}
\end{table}
\end{widetext}

We note that the coordinate transformation [Eq. (\ref{LK Hamiltonian}-\ref{mixing})] brings in the Euler angles $\alpha$, $\beta$, and $\gamma$ for process description. We note that the first two steps with $\alpha$ and $\beta$ angle involved already determine the finally-transformed $z$-axis, and the last step related to $\gamma$ angle only determines the $x$- and $y$-axis in the $x-y$ plane. Hence the intrinsic HH-LH mixing is irrelevant to the $\gamma$ angle. The intrinsic HH-LH mixing determined by QW orientations can be expressed as a function of the Euler angles $\alpha$ and $\beta$ as shown in Fig.~\ref{fig8}(b), where the correspondence between the Euler angles and QW orientations are listed in Table~\ref{tab2}.

Fig.~\ref{fig8} shows that the intrinsic HH-LH mixing remains zero in [001]-oriented QWs ($\beta=0$), but is maximal in [110]-oriented QWs ($\alpha=\pi/4$, $\beta=\pi/2$) with the value of
\begin{equation}\label{110-mixing}
C_{in}^{[110]}=\frac{\sqrt{3}}{2}\Big|\frac{\gamma_2-\gamma_3}{\gamma_1-2\gamma_2}\Big|.
\end{equation}
The zero intrinsic HH-LH mixing in [001]-oriented QWs implies that the symmetry-allowed HH-LH mixing completely originates from the local interface, despite the intrinsic HH-LH mixing in [110]-oriented QWs possesses the largest intrinsic HH-LH mixing due to its largest degree of the breaking of axial symmetry. Besides, the intrinsic HH-LH mixing in [111]-oriented QWs ($\alpha=\pi/4$, $\beta=\arccos\sqrt{3}/3$) remains zero, consistent with the vanishing $\bf{k}$-linear Rashba spin splitting in even-ML [111]-oriented QWs forbidden by the local interface symmetry. The emerging $\bf{k}$-linear Rashba spin splitting in odd-ML [111]-oriented QWs allowed by the local interface symmetry completely arises from the local-interface-induced HH-LH mixing rather than the intrinsic HH-LH mixing, same as the case of the [001]-oriented QWs. Along with the [$mmn$] crystalline orientations shown in Fig.~\ref{fig8}(c) [denoted as the dashed line in Fig.~\ref{fig8}(b)], the intrinsic HH-LH mixing drops from [110] (maximum) to [111] crystalline orientation (minimum, no mixing), rises from [111] to [112] crystalline orientation approximately, and decreases from [112] to [001] crystalline orientation (minimum, no mixing). Fig.~\ref{fig8}(c) shows a perfect consistency of the orientation-dependence, including the upper bound in [110] crystalline orientations. This consistency further confirms the intrinsic HH-LH mixing as the dominant factor regulating the relative strength relation of the $\bf{k}$-linear Rashba parameters.

\section{Summary}\label{section5}

In summary, we systematically study the orientation-dependent $\bf{k}$-linear or $\bf{k}$-cubic Rashba SOC of 2DHGs in QWs and find the upper bound to the $\bf{k}$-linear Rashba SOC in the [110] crystalline orientation and the purely $\bf{k}$-cubic Rashba SOC in the [111] orientation with an even number of the well MLs. Moreover, we show the tunability of the electric field and well width for both $\bf{k}$-linear and $\bf{k}$-cubic Rashba SOC. We analyze the presence or absence of the zone-center HH-LH mixing from the perspective of global and local interface symmetry, where the local interface symmetry governs the presence or absence of the $\bf{k}$-linear Rashba SOC and the global symmetry of the QW contributes to the enhancement of the $\bf{k}$-linear Rashba SOC. We illustrate that the intrinsic HH-LH mixing regulates the whole HH-LH mixing in low-symmetry orientations and the local-interface-induced HH-LH mixing is negligible in presence of the intrinsic counterpart unless the $\bf{k}$-linear Rashba SOC completely arises from the interface-induced HH-LH mixing for [001]-oriented and odd-ML [111]-oriented QWs. We directly prove that the postulation of the symmetry lowering of the local interface by observing the purely $\bf{k}$-cubic Rashba SOC in the even-ML [111]-oriented QWs in combination with the $\bf{k}$-linear Rashba SOC in the odd-ML [001]-oriented QWs. Furthermore, by doing the Hamiltonian transformation, we reveal the intrinsic HH-LH mixing regulating the relative strength relation of the $\bf{k}$-linear Rashba SOC in different QW orientations. These findings reveal the physical mechanism underlying orientation-dependence of the Rashba SOC in 2DHGs of QWs and provide strategic design principles to realize the linear or purely cubic Rashba SOC in 2D hole systems.

\begin{acknowledgements}
This work was supported by the National Science Fund for Distinguished Young Scholars under grant No. 11925407, the Basic Science Center Program of the National Natural Science Foundation of China (NSFC) under grant No. 61888102, and the Key Research Program of Frontier Sciences, CAS under grant No. ZDBS-LY-JSC019. S. G. was also supported by the NSFC under grant No. 11904359.
\end{acknowledgements}


\begin{thebibliography}{77}
\expandafter\ifx\csname natexlab\endcsname\relax\def\natexlab#1{#1}\fi
\expandafter\ifx\csname bibnamefont\endcsname\relax
  \def\bibnamefont#1{#1}\fi
\expandafter\ifx\csname bibfnamefont\endcsname\relax
  \def\bibfnamefont#1{#1}\fi
\expandafter\ifx\csname citenamefont\endcsname\relax
  \def\citenamefont#1{#1}\fi
\expandafter\ifx\csname url\endcsname\relax
  \def\url#1{\texttt{#1}}\fi
\expandafter\ifx\csname urlprefix\endcsname\relax\def\urlprefix{URL }\fi
\providecommand{\bibinfo}[2]{#2}
\providecommand{\eprint}[2][]{\url{#2}}

\bibitem[{\citenamefont{Dresselhaus}(1955)}]{Dresselhaus1955}
\bibinfo{author}{\bibfnamefont{G.}~\bibnamefont{Dresselhaus}},
  \bibinfo{journal}{Phys. Rev.} \textbf{\bibinfo{volume}{100}},
  \bibinfo{pages}{580} (\bibinfo{year}{1955}).

\bibitem[{\citenamefont{Bychkov and Rashba}(1984)}]{Rashba1984}
\bibinfo{author}{\bibfnamefont{Y.~A.} \bibnamefont{Bychkov}} \bibnamefont{and}
  \bibinfo{author}{\bibfnamefont{E.~I.} \bibnamefont{Rashba}},
  \bibinfo{journal}{JETP Lett.} \textbf{\bibinfo{volume}{39}},
  \bibinfo{pages}{78} (\bibinfo{year}{1984}).

\bibitem[{\citenamefont{Meier et~al.}(2007)\citenamefont{Meier, Salis,
  Shorubalko, Gini, Sch$\ddot{o}$n, and Ensslin}}]{Meier2007}
\bibinfo{author}{\bibfnamefont{L.}~\bibnamefont{Meier}},
  \bibinfo{author}{\bibfnamefont{G.}~\bibnamefont{Salis}},
  \bibinfo{author}{\bibfnamefont{I.}~\bibnamefont{Shorubalko}},
  \bibinfo{author}{\bibfnamefont{E.}~\bibnamefont{Gini}},
  \bibinfo{author}{\bibfnamefont{S.}~\bibnamefont{Sch$\ddot{o}$n}},
  \bibnamefont{and} \bibinfo{author}{\bibfnamefont{K.}~\bibnamefont{Ensslin}},
  \bibinfo{journal}{Nature Physics} \textbf{\bibinfo{volume}{3}},
  \bibinfo{pages}{650} (\bibinfo{year}{2007}).

\bibitem[{\citenamefont{Ishizaka et~al.}(2011)\citenamefont{Ishizaka, Bahramy,
  Murakawa, Sakano, Shimojima, Sonobe, Koizumi, Shin, Miyahara, Kimura
  et~al.}}]{Ishizaka_NM2011}
\bibinfo{author}{\bibfnamefont{K.}~\bibnamefont{Ishizaka}},
  \bibinfo{author}{\bibfnamefont{M.~S.} \bibnamefont{Bahramy}},
  \bibinfo{author}{\bibfnamefont{H.}~\bibnamefont{Murakawa}},
  \bibinfo{author}{\bibfnamefont{M.}~\bibnamefont{Sakano}},
  \bibinfo{author}{\bibfnamefont{T.}~\bibnamefont{Shimojima}},
  \bibinfo{author}{\bibfnamefont{T.}~\bibnamefont{Sonobe}},
  \bibinfo{author}{\bibfnamefont{K.}~\bibnamefont{Koizumi}},
  \bibinfo{author}{\bibfnamefont{S.}~\bibnamefont{Shin}},
  \bibinfo{author}{\bibfnamefont{H.}~\bibnamefont{Miyahara}},
  \bibinfo{author}{\bibfnamefont{A.}~\bibnamefont{Kimura}},
  \bibnamefont{et~al.}, \bibinfo{journal}{Nat. Mater.}
  \textbf{\bibinfo{volume}{10}}, \bibinfo{pages}{521} (\bibinfo{year}{2011}).

\bibitem[{\citenamefont{Zhang et~al.}(2014)\citenamefont{Zhang, Liu, Luo,
  Freeman, and Zunger}}]{LuoNP2014}
\bibinfo{author}{\bibfnamefont{X.-W.} \bibnamefont{Zhang}},
  \bibinfo{author}{\bibfnamefont{Q.-H.} \bibnamefont{Liu}},
  \bibinfo{author}{\bibfnamefont{J.-W.} \bibnamefont{Luo}},
  \bibinfo{author}{\bibfnamefont{A.~J.} \bibnamefont{Freeman}},
  \bibnamefont{and} \bibinfo{author}{\bibfnamefont{A.}~\bibnamefont{Zunger}},
  \bibinfo{journal}{Nat. Phys.} \textbf{\bibinfo{volume}{10}},
  \bibinfo{pages}{387} (\bibinfo{year}{2014}).

\bibitem[{\citenamefont{Manchon et~al.}(2015)\citenamefont{Manchon, Koo, Nitta,
  Frolov, and Duine}}]{Manchon2015}
\bibinfo{author}{\bibfnamefont{A.}~\bibnamefont{Manchon}},
  \bibinfo{author}{\bibfnamefont{H.~C.} \bibnamefont{Koo}},
  \bibinfo{author}{\bibfnamefont{J.}~\bibnamefont{Nitta}},
  \bibinfo{author}{\bibfnamefont{S.~M.} \bibnamefont{Frolov}},
  \bibnamefont{and} \bibinfo{author}{\bibfnamefont{R.~A.} \bibnamefont{Duine}},
  \bibinfo{journal}{Nat. Mater.} \textbf{\bibinfo{volume}{14}},
  \bibinfo{pages}{871} (\bibinfo{year}{2015}).

\bibitem[{\citenamefont{Sinova et~al.}(2004)\citenamefont{Sinova, Culcer, Niu,
  Sinitsyn, Jungwirth, and MacDonald}}]{Sinova2004}
\bibinfo{author}{\bibfnamefont{J.}~\bibnamefont{Sinova}},
  \bibinfo{author}{\bibfnamefont{D.}~\bibnamefont{Culcer}},
  \bibinfo{author}{\bibfnamefont{Q.}~\bibnamefont{Niu}},
  \bibinfo{author}{\bibfnamefont{N.~A.} \bibnamefont{Sinitsyn}},
  \bibinfo{author}{\bibfnamefont{T.}~\bibnamefont{Jungwirth}},
  \bibnamefont{and} \bibinfo{author}{\bibfnamefont{A.~H.}
  \bibnamefont{MacDonald}}, \bibinfo{journal}{Phys. Rev. Lett.}
  \textbf{\bibinfo{volume}{92}}, \bibinfo{pages}{126603}
  (\bibinfo{year}{2004}).

\bibitem[{\citenamefont{Bernevig and Zhang}(2005)}]{Bernevig2005}
\bibinfo{author}{\bibfnamefont{B.~A.} \bibnamefont{Bernevig}} \bibnamefont{and}
  \bibinfo{author}{\bibfnamefont{S.-C.} \bibnamefont{Zhang}},
  \bibinfo{journal}{Phys. Rev. Lett.} \textbf{\bibinfo{volume}{95}},
  \bibinfo{pages}{016801} (\bibinfo{year}{2005}).

\bibitem[{\citenamefont{Kato et~al.}(2004)\citenamefont{Kato, Myers, Gossard,
  and Awschalom}}]{Awschalom2004}
\bibinfo{author}{\bibfnamefont{Y.~K.} \bibnamefont{Kato}},
  \bibinfo{author}{\bibfnamefont{R.~C.} \bibnamefont{Myers}},
  \bibinfo{author}{\bibfnamefont{A.~C.} \bibnamefont{Gossard}},
  \bibnamefont{and} \bibinfo{author}{\bibfnamefont{D.~D.}
  \bibnamefont{Awschalom}}, \bibinfo{journal}{Science}
  \textbf{\bibinfo{volume}{306}}, \bibinfo{pages}{1910} (\bibinfo{year}{2004}).

\bibitem[{\citenamefont{Wunderlich et~al.}(2009)\citenamefont{Wunderlich,
  Irvine, Sinova, Park, Zarbo, Xu, Kaestner, Novak, and
  Jungwirth}}]{Wunderlich2009}
\bibinfo{author}{\bibfnamefont{J.}~\bibnamefont{Wunderlich}},
  \bibinfo{author}{\bibfnamefont{A.~C.} \bibnamefont{Irvine}},
  \bibinfo{author}{\bibfnamefont{J.}~\bibnamefont{Sinova}},
  \bibinfo{author}{\bibfnamefont{B.~G.} \bibnamefont{Park}},
  \bibinfo{author}{\bibfnamefont{L.~P.} \bibnamefont{Zarbo}},
  \bibinfo{author}{\bibfnamefont{X.~L.} \bibnamefont{Xu}},
  \bibinfo{author}{\bibfnamefont{B.}~\bibnamefont{Kaestner}},
  \bibinfo{author}{\bibfnamefont{V.}~\bibnamefont{Novak}}, \bibnamefont{and}
  \bibinfo{author}{\bibfnamefont{T.}~\bibnamefont{Jungwirth}},
  \bibinfo{journal}{Nature Physics} \textbf{\bibinfo{volume}{5}},
  \bibinfo{pages}{675} (\bibinfo{year}{2009}).

\bibitem[{\citenamefont{Ivchenko and Pikus}(1978)}]{Ivchenko1978}
\bibinfo{author}{\bibfnamefont{G.~E.} \bibnamefont{Ivchenko},
  \bibfnamefont{E.~L.}} \bibnamefont{and}
  \bibinfo{author}{\bibnamefont{Pikus}}, \bibinfo{journal}{JETP Lett.}
  \textbf{\bibinfo{volume}{27}}, \bibinfo{pages}{604} (\bibinfo{year}{1978}).

\bibitem[{\citenamefont{Ganichev}(2008)}]{Ganichev2008}
\bibinfo{author}{\bibfnamefont{S.~D.} \bibnamefont{Ganichev}},
  \bibinfo{journal}{Int. J. Mod. Phys. B} \textbf{\bibinfo{volume}{22}},
  \bibinfo{pages}{1} (\bibinfo{year}{2008}).

\bibitem[{\citenamefont{Datta and Das}(1990)}]{Datta1990}
\bibinfo{author}{\bibfnamefont{S.}~\bibnamefont{Datta}} \bibnamefont{and}
  \bibinfo{author}{\bibfnamefont{B.}~\bibnamefont{Das}},
  \bibinfo{journal}{Appl. Phys. Lett.} \textbf{\bibinfo{volume}{56}},
  \bibinfo{pages}{665} (\bibinfo{year}{1990}).

\bibitem[{\citenamefont{Schliemann et~al.}(2003)\citenamefont{Schliemann,
  Egues, and Loss}}]{Schliemann2003}
\bibinfo{author}{\bibfnamefont{J.}~\bibnamefont{Schliemann}},
  \bibinfo{author}{\bibfnamefont{J.~C.} \bibnamefont{Egues}}, \bibnamefont{and}
  \bibinfo{author}{\bibfnamefont{D.}~\bibnamefont{Loss}},
  \bibinfo{journal}{Phys. Rev. Lett.} \textbf{\bibinfo{volume}{90}},
  \bibinfo{pages}{146801} (\bibinfo{year}{2003}).

\bibitem[{\citenamefont{Maurand et~al.}(2016)\citenamefont{Maurand, Jehl,
  Kotekar-Patil, Corna, Bohuslavskyi, Lavieville, Hutin, Barraud, Vinet,
  Sanquer et~al.}}]{Maurand2016}
\bibinfo{author}{\bibfnamefont{R.}~\bibnamefont{Maurand}},
  \bibinfo{author}{\bibfnamefont{X.}~\bibnamefont{Jehl}},
  \bibinfo{author}{\bibfnamefont{D.}~\bibnamefont{Kotekar-Patil}},
  \bibinfo{author}{\bibfnamefont{A.}~\bibnamefont{Corna}},
  \bibinfo{author}{\bibfnamefont{H.}~\bibnamefont{Bohuslavskyi}},
  \bibinfo{author}{\bibfnamefont{R.}~\bibnamefont{Lavieville}},
  \bibinfo{author}{\bibfnamefont{L.}~\bibnamefont{Hutin}},
  \bibinfo{author}{\bibfnamefont{S.}~\bibnamefont{Barraud}},
  \bibinfo{author}{\bibfnamefont{M.}~\bibnamefont{Vinet}},
  \bibinfo{author}{\bibfnamefont{M.}~\bibnamefont{Sanquer}},
  \bibnamefont{et~al.}, \bibinfo{journal}{Nat. Commun.}
  \textbf{\bibinfo{volume}{7}}, \bibinfo{pages}{13575} (\bibinfo{year}{2016}).

\bibitem[{\citenamefont{Watzinger et~al.}(2018)\citenamefont{Watzinger,
  Kukucka, Vukusic, Gao, Wang, Schaffler, Zhang, and Katsaros}}]{Watzinger2018}
\bibinfo{author}{\bibfnamefont{H.}~\bibnamefont{Watzinger}},
  \bibinfo{author}{\bibfnamefont{J.}~\bibnamefont{Kukucka}},
  \bibinfo{author}{\bibfnamefont{L.}~\bibnamefont{Vukusic}},
  \bibinfo{author}{\bibfnamefont{F.}~\bibnamefont{Gao}},
  \bibinfo{author}{\bibfnamefont{T.}~\bibnamefont{Wang}},
  \bibinfo{author}{\bibfnamefont{F.}~\bibnamefont{Schaffler}},
  \bibinfo{author}{\bibfnamefont{J.-J.} \bibnamefont{Zhang}}, \bibnamefont{and}
  \bibinfo{author}{\bibfnamefont{G.}~\bibnamefont{Katsaros}},
  \bibinfo{journal}{Nature Communications} \textbf{\bibinfo{volume}{9}},
  \bibinfo{pages}{3902} (\bibinfo{year}{2018}).

\bibitem[{\citenamefont{Hendrickx et~al.}(2018)\citenamefont{Hendrickx, Franke,
  Sammak, Kouwenhoven, Sabbagh, Yeoh, Li, Tagliaferri, Virgilio, Capellini
  et~al.}}]{Hendrickx2018}
\bibinfo{author}{\bibfnamefont{N.~W.} \bibnamefont{Hendrickx}},
  \bibinfo{author}{\bibfnamefont{D.~P.} \bibnamefont{Franke}},
  \bibinfo{author}{\bibfnamefont{A.}~\bibnamefont{Sammak}},
  \bibinfo{author}{\bibfnamefont{M.}~\bibnamefont{Kouwenhoven}},
  \bibinfo{author}{\bibfnamefont{D.}~\bibnamefont{Sabbagh}},
  \bibinfo{author}{\bibfnamefont{L.}~\bibnamefont{Yeoh}},
  \bibinfo{author}{\bibfnamefont{R.}~\bibnamefont{Li}},
  \bibinfo{author}{\bibfnamefont{M.~L.~V.} \bibnamefont{Tagliaferri}},
  \bibinfo{author}{\bibfnamefont{M.}~\bibnamefont{Virgilio}},
  \bibinfo{author}{\bibfnamefont{G.}~\bibnamefont{Capellini}},
  \bibnamefont{et~al.}, \bibinfo{journal}{Nat. Commun.}
  \textbf{\bibinfo{volume}{9}}, \bibinfo{pages}{2835} (\bibinfo{year}{2018}).

\bibitem[{\citenamefont{Brauns et~al.}(2016)\citenamefont{Brauns, Ridderbos,
  Li, van~der Wiel, Bakkers, and Zwanenburg}}]{Brauns2016}
\bibinfo{author}{\bibfnamefont{M.}~\bibnamefont{Brauns}},
  \bibinfo{author}{\bibfnamefont{J.}~\bibnamefont{Ridderbos}},
  \bibinfo{author}{\bibfnamefont{A.}~\bibnamefont{Li}},
  \bibinfo{author}{\bibfnamefont{W.~G.} \bibnamefont{van~der Wiel}},
  \bibinfo{author}{\bibfnamefont{E.~P. A.~M.} \bibnamefont{Bakkers}},
  \bibnamefont{and} \bibinfo{author}{\bibfnamefont{F.~A.}
  \bibnamefont{Zwanenburg}}, \bibinfo{journal}{Appl. Phys. Lett.}
  \textbf{\bibinfo{volume}{109}}, \bibinfo{pages}{143113}
  (\bibinfo{year}{2016}).

\bibitem[{\citenamefont{Tyryshkin et~al.}(2012)\citenamefont{Tyryshkin, Tojo,
  Morton, Riemann, Abrosimov, Becker, Pohl, Schenkel, Thewalt, Itoh
  et~al.}}]{Tyryshkin2012}
\bibinfo{author}{\bibfnamefont{A.~M.} \bibnamefont{Tyryshkin}},
  \bibinfo{author}{\bibfnamefont{S.}~\bibnamefont{Tojo}},
  \bibinfo{author}{\bibfnamefont{J.~J.~L.} \bibnamefont{Morton}},
  \bibinfo{author}{\bibfnamefont{H.}~\bibnamefont{Riemann}},
  \bibinfo{author}{\bibfnamefont{N.~V.} \bibnamefont{Abrosimov}},
  \bibinfo{author}{\bibfnamefont{P.}~\bibnamefont{Becker}},
  \bibinfo{author}{\bibfnamefont{H.-J.} \bibnamefont{Pohl}},
  \bibinfo{author}{\bibfnamefont{T.}~\bibnamefont{Schenkel}},
  \bibinfo{author}{\bibfnamefont{M.~L.~W.} \bibnamefont{Thewalt}},
  \bibinfo{author}{\bibfnamefont{K.~M.} \bibnamefont{Itoh}},
  \bibnamefont{et~al.}, \bibinfo{journal}{Nature Materials}
  \textbf{\bibinfo{volume}{11}}, \bibinfo{pages}{143} (\bibinfo{year}{2012}).

\bibitem[{\citenamefont{Veldhorst et~al.}(2014)\citenamefont{Veldhorst, Hwang,
  Yang, Leenstra, de~Ronde, Dehollain, Muhonen, Hudson, Itoh, Morello
  et~al.}}]{Veldhorst2014}
\bibinfo{author}{\bibfnamefont{M.}~\bibnamefont{Veldhorst}},
  \bibinfo{author}{\bibfnamefont{J.~C.~C.} \bibnamefont{Hwang}},
  \bibinfo{author}{\bibfnamefont{C.~H.} \bibnamefont{Yang}},
  \bibinfo{author}{\bibfnamefont{A.~W.} \bibnamefont{Leenstra}},
  \bibinfo{author}{\bibfnamefont{B.}~\bibnamefont{de~Ronde}},
  \bibinfo{author}{\bibfnamefont{J.~P.} \bibnamefont{Dehollain}},
  \bibinfo{author}{\bibfnamefont{J.~T.} \bibnamefont{Muhonen}},
  \bibinfo{author}{\bibfnamefont{F.~E.} \bibnamefont{Hudson}},
  \bibinfo{author}{\bibfnamefont{K.~M.} \bibnamefont{Itoh}},
  \bibinfo{author}{\bibfnamefont{A.}~\bibnamefont{Morello}},
  \bibnamefont{et~al.}, \bibinfo{journal}{Nature Nanotechnology}
  \textbf{\bibinfo{volume}{9}}, \bibinfo{pages}{981} (\bibinfo{year}{2014}).

\bibitem[{\citenamefont{Yoneda et~al.}(2018)\citenamefont{Yoneda, Takeda,
  Otsuka, Nakajima, Delbecq, Allison, Honda, Kodera, Oda, Hoshi
  et~al.}}]{Yoneda2018}
\bibinfo{author}{\bibfnamefont{J.}~\bibnamefont{Yoneda}},
  \bibinfo{author}{\bibfnamefont{K.}~\bibnamefont{Takeda}},
  \bibinfo{author}{\bibfnamefont{T.}~\bibnamefont{Otsuka}},
  \bibinfo{author}{\bibfnamefont{T.}~\bibnamefont{Nakajima}},
  \bibinfo{author}{\bibfnamefont{M.~R.} \bibnamefont{Delbecq}},
  \bibinfo{author}{\bibfnamefont{G.}~\bibnamefont{Allison}},
  \bibinfo{author}{\bibfnamefont{T.}~\bibnamefont{Honda}},
  \bibinfo{author}{\bibfnamefont{T.}~\bibnamefont{Kodera}},
  \bibinfo{author}{\bibfnamefont{S.}~\bibnamefont{Oda}},
  \bibinfo{author}{\bibfnamefont{Y.}~\bibnamefont{Hoshi}},
  \bibnamefont{et~al.}, \bibinfo{journal}{Nature Nanotechnology}
  \textbf{\bibinfo{volume}{13}}, \bibinfo{pages}{102} (\bibinfo{year}{2018}).

\bibitem[{\citenamefont{Hu et~al.}(2011)\citenamefont{Hu, Kuemmeth, Lieber, and
  Marcus}}]{Hu2011}
\bibinfo{author}{\bibfnamefont{Y.}~\bibnamefont{Hu}},
  \bibinfo{author}{\bibfnamefont{F.}~\bibnamefont{Kuemmeth}},
  \bibinfo{author}{\bibfnamefont{C.~M.} \bibnamefont{Lieber}},
  \bibnamefont{and} \bibinfo{author}{\bibfnamefont{C.~M.}
  \bibnamefont{Marcus}}, \bibinfo{journal}{Nat. Nanotechnol.}
  \textbf{\bibinfo{volume}{7}}, \bibinfo{pages}{47} (\bibinfo{year}{2011}).

\bibitem[{\citenamefont{Warburton}(2013)}]{Warburton2013}
\bibinfo{author}{\bibfnamefont{R.~J.} \bibnamefont{Warburton}},
  \bibinfo{journal}{Nat. Mater.} \textbf{\bibinfo{volume}{12}},
  \bibinfo{pages}{483} (\bibinfo{year}{2013}).

\bibitem[{\citenamefont{Hendrickx et~al.}(2020)\citenamefont{Hendrickx, Franke,
  Sammak, Scappucci, and Veldhorst}}]{Hendrickx2020}
\bibinfo{author}{\bibfnamefont{N.~W.} \bibnamefont{Hendrickx}},
  \bibinfo{author}{\bibfnamefont{D.~P.} \bibnamefont{Franke}},
  \bibinfo{author}{\bibfnamefont{A.}~\bibnamefont{Sammak}},
  \bibinfo{author}{\bibfnamefont{G.}~\bibnamefont{Scappucci}},
  \bibnamefont{and}
  \bibinfo{author}{\bibfnamefont{M.}~\bibnamefont{Veldhorst}},
  \bibinfo{journal}{Nature} \textbf{\bibinfo{volume}{577}},
  \bibinfo{pages}{487} (\bibinfo{year}{2020}).

\bibitem[{\citenamefont{Hendrickx et~al.}(2021)\citenamefont{Hendrickx, Lawrie,
  Russ, van Riggelen, de~Snoo, Schouten, Sammak, Scappucci, and
  Veldhorst}}]{Hendrickx2021}
\bibinfo{author}{\bibfnamefont{N.~W.} \bibnamefont{Hendrickx}},
  \bibinfo{author}{\bibfnamefont{W.~I.~L.} \bibnamefont{Lawrie}},
  \bibinfo{author}{\bibfnamefont{M.}~\bibnamefont{Russ}},
  \bibinfo{author}{\bibfnamefont{F.}~\bibnamefont{van Riggelen}},
  \bibinfo{author}{\bibfnamefont{S.~L.} \bibnamefont{de~Snoo}},
  \bibinfo{author}{\bibfnamefont{R.~N.} \bibnamefont{Schouten}},
  \bibinfo{author}{\bibfnamefont{A.}~\bibnamefont{Sammak}},
  \bibinfo{author}{\bibfnamefont{G.}~\bibnamefont{Scappucci}},
  \bibnamefont{and}
  \bibinfo{author}{\bibfnamefont{M.}~\bibnamefont{Veldhorst}},
  \bibinfo{journal}{Nature} \textbf{\bibinfo{volume}{591}},
  \bibinfo{pages}{580} (\bibinfo{year}{2021}).

\bibitem[{\citenamefont{Sammak et~al.}(2021)\citenamefont{Sammak, Sabbagh,
  Hendrickx, Lodari, Paquelet~Wuetz, Tosato, Yeoh, Bollani, Virgilio, Schubert
  et~al.}}]{Sammak2021}
\bibinfo{author}{\bibfnamefont{A.}~\bibnamefont{Sammak}},
  \bibinfo{author}{\bibfnamefont{D.}~\bibnamefont{Sabbagh}},
  \bibinfo{author}{\bibfnamefont{N.~W.} \bibnamefont{Hendrickx}},
  \bibinfo{author}{\bibfnamefont{M.}~\bibnamefont{Lodari}},
  \bibinfo{author}{\bibfnamefont{B.}~\bibnamefont{Paquelet~Wuetz}},
  \bibinfo{author}{\bibfnamefont{A.}~\bibnamefont{Tosato}},
  \bibinfo{author}{\bibfnamefont{L.}~\bibnamefont{Yeoh}},
  \bibinfo{author}{\bibfnamefont{M.}~\bibnamefont{Bollani}},
  \bibinfo{author}{\bibfnamefont{M.}~\bibnamefont{Virgilio}},
  \bibinfo{author}{\bibfnamefont{M.~A.} \bibnamefont{Schubert}},
  \bibnamefont{et~al.}, \bibinfo{journal}{Adv. Funct. Mater.}
  \textbf{\bibinfo{volume}{29}}, \bibinfo{pages}{1807613}
  (\bibinfo{year}{2021}).

\bibitem[{\citenamefont{Bulaev and Loss}(2007)}]{EDSR2007}
\bibinfo{author}{\bibfnamefont{D.~V.} \bibnamefont{Bulaev}} \bibnamefont{and}
  \bibinfo{author}{\bibfnamefont{D.}~\bibnamefont{Loss}},
  \bibinfo{journal}{Phys. Rev. Lett.} \textbf{\bibinfo{volume}{98}},
  \bibinfo{pages}{097202} (\bibinfo{year}{2007}).

\bibitem[{\citenamefont{Bulaev and Loss}(2005)}]{Bulaev2005}
\bibinfo{author}{\bibfnamefont{D.~V.} \bibnamefont{Bulaev}} \bibnamefont{and}
  \bibinfo{author}{\bibfnamefont{D.}~\bibnamefont{Loss}},
  \bibinfo{journal}{Phys. Rev. Lett.} \textbf{\bibinfo{volume}{95}},
  \bibinfo{pages}{076805} (\bibinfo{year}{2005}).

\bibitem[{\citenamefont{Scappucci et~al.}(2020)\citenamefont{Scappucci,
  Kloeffel, Zwanenburg, Loss, Myronov, Zhang, De~Franceschi, Katsaros, and
  Veldhorst}}]{Scappucci2020}
\bibinfo{author}{\bibfnamefont{G.}~\bibnamefont{Scappucci}},
  \bibinfo{author}{\bibfnamefont{C.}~\bibnamefont{Kloeffel}},
  \bibinfo{author}{\bibfnamefont{F.~A.} \bibnamefont{Zwanenburg}},
  \bibinfo{author}{\bibfnamefont{D.}~\bibnamefont{Loss}},
  \bibinfo{author}{\bibfnamefont{M.}~\bibnamefont{Myronov}},
  \bibinfo{author}{\bibfnamefont{J.-J.} \bibnamefont{Zhang}},
  \bibinfo{author}{\bibfnamefont{S.}~\bibnamefont{De~Franceschi}},
  \bibinfo{author}{\bibfnamefont{G.}~\bibnamefont{Katsaros}}, \bibnamefont{and}
  \bibinfo{author}{\bibfnamefont{M.}~\bibnamefont{Veldhorst}},
  \bibinfo{journal}{Nature Reviews Materials}  (\bibinfo{year}{2020}).

\bibitem[{\citenamefont{Luo et~al.}(2011)\citenamefont{Luo, Zhang, and
  Zunger}}]{Luo2011}
\bibinfo{author}{\bibfnamefont{J.-W.} \bibnamefont{Luo}},
  \bibinfo{author}{\bibfnamefont{L.}~\bibnamefont{Zhang}}, \bibnamefont{and}
  \bibinfo{author}{\bibfnamefont{A.}~\bibnamefont{Zunger}},
  \bibinfo{journal}{Phys. Rev. B} \textbf{\bibinfo{volume}{84}},
  \bibinfo{pages}{121303(R)} (\bibinfo{year}{2011}).

\bibitem[{\citenamefont{Xiong et~al.}(2021)\citenamefont{Xiong, Guan, Luo, and
  Li}}]{Xiong_PRB2021}
\bibinfo{author}{\bibfnamefont{J.-X.} \bibnamefont{Xiong}},
  \bibinfo{author}{\bibfnamefont{S.}~\bibnamefont{Guan}},
  \bibinfo{author}{\bibfnamefont{J.-W.} \bibnamefont{Luo}}, \bibnamefont{and}
  \bibinfo{author}{\bibfnamefont{S.-S.} \bibnamefont{Li}},
  \bibinfo{journal}{Phys. Rev. B} \textbf{\bibinfo{volume}{103}},
  \bibinfo{pages}{085309} (\bibinfo{year}{2021}).

\bibitem[{\citenamefont{Moriya et~al.}(2014)\citenamefont{Moriya, Sawano,
  Hoshi, Masubuchi, Shiraki, Wild, Neumann, Abstreiter, Bougeard, Koga
  et~al.}}]{Moriya2014}
\bibinfo{author}{\bibfnamefont{R.}~\bibnamefont{Moriya}},
  \bibinfo{author}{\bibfnamefont{K.}~\bibnamefont{Sawano}},
  \bibinfo{author}{\bibfnamefont{Y.}~\bibnamefont{Hoshi}},
  \bibinfo{author}{\bibfnamefont{S.}~\bibnamefont{Masubuchi}},
  \bibinfo{author}{\bibfnamefont{Y.}~\bibnamefont{Shiraki}},
  \bibinfo{author}{\bibfnamefont{A.}~\bibnamefont{Wild}},
  \bibinfo{author}{\bibfnamefont{C.}~\bibnamefont{Neumann}},
  \bibinfo{author}{\bibfnamefont{G.}~\bibnamefont{Abstreiter}},
  \bibinfo{author}{\bibfnamefont{D.}~\bibnamefont{Bougeard}},
  \bibinfo{author}{\bibfnamefont{T.}~\bibnamefont{Koga}}, \bibnamefont{et~al.},
  \bibinfo{journal}{Phys. Rev. Lett.} \textbf{\bibinfo{volume}{113}},
  \bibinfo{pages}{086601} (\bibinfo{year}{2014}).

\bibitem[{\citenamefont{Winkler}(2003)}]{Winkler2003}
\bibinfo{author}{\bibfnamefont{R.}~\bibnamefont{Winkler}},
  \bibinfo{title}{Spin-Orbit Coupling Effect in Two-Dimensional Electron
  and Hole Systems} (\bibinfo{publisher}{Springer}, \bibinfo{year}{2003}).

\bibitem[{\citenamefont{Kloeffel et~al.}(2018)\citenamefont{Kloeffel,
  Ran\ifmmode \check{c}\else \v{c}\fi{}i\ifmmode~\acute{c}\else \'{c}\fi{}, and
  Loss}}]{Kloeffel2018a}
\bibinfo{author}{\bibfnamefont{C.}~\bibnamefont{Kloeffel}},
  \bibinfo{author}{\bibfnamefont{M.~J.} \bibnamefont{Ran\ifmmode \check{c}\else
  \v{c}\fi{}i\ifmmode~\acute{c}\else \'{c}\fi{}}}, \bibnamefont{and}
  \bibinfo{author}{\bibfnamefont{D.}~\bibnamefont{Loss}},
  \bibinfo{journal}{Phys. Rev. B} \textbf{\bibinfo{volume}{97}},
  \bibinfo{pages}{235422} (\bibinfo{year}{2018}).

\bibitem[{\citenamefont{Schliemann and Loss}(2005)}]{Schliemann2005}
\bibinfo{author}{\bibfnamefont{J.}~\bibnamefont{Schliemann}} \bibnamefont{and}
  \bibinfo{author}{\bibfnamefont{D.}~\bibnamefont{Loss}},
  \bibinfo{journal}{Phys. Rev. B} \textbf{\bibinfo{volume}{71}},
  \bibinfo{pages}{085308} (\bibinfo{year}{2005}).

\bibitem[{\citenamefont{Failla et~al.}(2015)\citenamefont{Failla, Myronov,
  Morrison, Leadley, and Lloyd-Hughes}}]{Failla_PRB2015}
\bibinfo{author}{\bibfnamefont{M.}~\bibnamefont{Failla}},
  \bibinfo{author}{\bibfnamefont{M.}~\bibnamefont{Myronov}},
  \bibinfo{author}{\bibfnamefont{C.}~\bibnamefont{Morrison}},
  \bibinfo{author}{\bibfnamefont{D.~R.} \bibnamefont{Leadley}},
  \bibnamefont{and}
  \bibinfo{author}{\bibfnamefont{J.}~\bibnamefont{Lloyd-Hughes}},
  \bibinfo{journal}{Phys. Rev. B} \textbf{\bibinfo{volume}{92}},
  \bibinfo{pages}{045303} (\bibinfo{year}{2015}).

\bibitem[{\citenamefont{Nakamura et~al.}(2012)\citenamefont{Nakamura, Koga, and
  Kimura}}]{Nakamura2012}
\bibinfo{author}{\bibfnamefont{H.}~\bibnamefont{Nakamura}},
  \bibinfo{author}{\bibfnamefont{T.}~\bibnamefont{Koga}}, \bibnamefont{and}
  \bibinfo{author}{\bibfnamefont{T.}~\bibnamefont{Kimura}},
  \bibinfo{journal}{Phys. Rev. Lett.} \textbf{\bibinfo{volume}{108}},
  \bibinfo{pages}{206601} (\bibinfo{year}{2012}).

\bibitem[{\citenamefont{Shanavas}(2016)}]{Shanavas2016}
\bibinfo{author}{\bibfnamefont{K.~V.} \bibnamefont{Shanavas}},
  \bibinfo{journal}{Phys. Rev. B} \textbf{\bibinfo{volume}{93}},
  \bibinfo{pages}{045108} (\bibinfo{year}{2016}).

\bibitem[{\citenamefont{Lin et~al.}(2019)\citenamefont{Lin, Li,
  Do$\check{g}$an, Li, Rotella, Yu, Zhang, Li, Lew, Wang et~al.}}]{Lin2019}
\bibinfo{author}{\bibfnamefont{W.}~\bibnamefont{Lin}},
  \bibinfo{author}{\bibfnamefont{L.}~\bibnamefont{Li}},
  \bibinfo{author}{\bibfnamefont{F.}~\bibnamefont{Do$\check{g}$an}},
  \bibinfo{author}{\bibfnamefont{C.}~\bibnamefont{Li}},
  \bibinfo{author}{\bibfnamefont{H.}~\bibnamefont{Rotella}},
  \bibinfo{author}{\bibfnamefont{X.}~\bibnamefont{Yu}},
  \bibinfo{author}{\bibfnamefont{B.}~\bibnamefont{Zhang}},
  \bibinfo{author}{\bibfnamefont{Y.}~\bibnamefont{Li}},
  \bibinfo{author}{\bibfnamefont{W.~S.} \bibnamefont{Lew}},
  \bibinfo{author}{\bibfnamefont{S.}~\bibnamefont{Wang}}, \bibnamefont{et~al.},
  \bibinfo{journal}{Nature Communications} \textbf{\bibinfo{volume}{10}},
  \bibinfo{pages}{3052} (\bibinfo{year}{2019}).

\bibitem[{\citenamefont{Usachov et~al.}(2020)\citenamefont{Usachov, Nechaev,
  Poelchen, G\"uttler, Krasovskii, Schulz, Generalov, Kliemt, Kraiker, Krellner
  et~al.}}]{Usachov_PRL2020}
\bibinfo{author}{\bibfnamefont{D.~Y.} \bibnamefont{Usachov}},
  \bibinfo{author}{\bibfnamefont{I.~A.} \bibnamefont{Nechaev}},
  \bibinfo{author}{\bibfnamefont{G.}~\bibnamefont{Poelchen}},
  \bibinfo{author}{\bibfnamefont{M.}~\bibnamefont{G\"uttler}},
  \bibinfo{author}{\bibfnamefont{E.~E.} \bibnamefont{Krasovskii}},
  \bibinfo{author}{\bibfnamefont{S.}~\bibnamefont{Schulz}},
  \bibinfo{author}{\bibfnamefont{A.}~\bibnamefont{Generalov}},
  \bibinfo{author}{\bibfnamefont{K.}~\bibnamefont{Kliemt}},
  \bibinfo{author}{\bibfnamefont{A.}~\bibnamefont{Kraiker}},
  \bibinfo{author}{\bibfnamefont{C.}~\bibnamefont{Krellner}},
  \bibnamefont{et~al.}, \bibinfo{journal}{Phys. Rev. Lett.}
  \textbf{\bibinfo{volume}{124}}, \bibinfo{pages}{237202}
  (\bibinfo{year}{2020}).

\bibitem[{\citenamefont{Chesi et~al.}(2011)\citenamefont{Chesi, Giuliani,
  Rokhinson, Pfeiffer, and West}}]{Chesi_PRL2011}
\bibinfo{author}{\bibfnamefont{S.}~\bibnamefont{Chesi}},
  \bibinfo{author}{\bibfnamefont{G.~F.} \bibnamefont{Giuliani}},
  \bibinfo{author}{\bibfnamefont{L.~P.} \bibnamefont{Rokhinson}},
  \bibinfo{author}{\bibfnamefont{L.~N.} \bibnamefont{Pfeiffer}},
  \bibnamefont{and} \bibinfo{author}{\bibfnamefont{K.~W.} \bibnamefont{West}},
  \bibinfo{journal}{Phys. Rev. Lett.} \textbf{\bibinfo{volume}{106}},
  \bibinfo{pages}{236601} (\bibinfo{year}{2011}).

\bibitem[{\citenamefont{Zhao et~al.}(2020)\citenamefont{Zhao, Nakamura, Arras,
  Paillard, Chen, Gosteau, Li, Yang, and Bellaiche}}]{Zhao_PRL2020}
\bibinfo{author}{\bibfnamefont{H.~J.} \bibnamefont{Zhao}},
  \bibinfo{author}{\bibfnamefont{H.}~\bibnamefont{Nakamura}},
  \bibinfo{author}{\bibfnamefont{R.}~\bibnamefont{Arras}},
  \bibinfo{author}{\bibfnamefont{C.}~\bibnamefont{Paillard}},
  \bibinfo{author}{\bibfnamefont{P.}~\bibnamefont{Chen}},
  \bibinfo{author}{\bibfnamefont{J.}~\bibnamefont{Gosteau}},
  \bibinfo{author}{\bibfnamefont{X.}~\bibnamefont{Li}},
  \bibinfo{author}{\bibfnamefont{Y.}~\bibnamefont{Yang}}, \bibnamefont{and}
  \bibinfo{author}{\bibfnamefont{L.}~\bibnamefont{Bellaiche}},
  \bibinfo{journal}{Phys. Rev. Lett.} \textbf{\bibinfo{volume}{125}},
  \bibinfo{pages}{216405} (\bibinfo{year}{2020}).

\bibitem[{\citenamefont{Wang and Zunger}(1994{\natexlab{a}})}]{Wang1994}
\bibinfo{author}{\bibfnamefont{L.-W.} \bibnamefont{Wang}} \bibnamefont{and}
  \bibinfo{author}{\bibfnamefont{A.}~\bibnamefont{Zunger}},
  \bibinfo{journal}{J. Chem. Phys.} \textbf{\bibinfo{volume}{100}},
  \bibinfo{pages}{2394} (\bibinfo{year}{1994}{\natexlab{a}}).

\bibitem[{\citenamefont{Wang and Zunger}(1995)}]{Wang1995}
\bibinfo{author}{\bibfnamefont{L.-W.} \bibnamefont{Wang}} \bibnamefont{and}
  \bibinfo{author}{\bibfnamefont{A.}~\bibnamefont{Zunger}},
  \bibinfo{journal}{Phys. Rev. B} \textbf{\bibinfo{volume}{51}},
  \bibinfo{pages}{17398} (\bibinfo{year}{1995}).

\bibitem[{\citenamefont{Wang et~al.}(1999)\citenamefont{Wang, Kim, and
  Zunger}}]{Wang1999}
\bibinfo{author}{\bibfnamefont{L.-W.} \bibnamefont{Wang}},
  \bibinfo{author}{\bibfnamefont{J.}~\bibnamefont{Kim}}, \bibnamefont{and}
  \bibinfo{author}{\bibfnamefont{A.}~\bibnamefont{Zunger}},
  \bibinfo{journal}{Phys. Rev. B} \textbf{\bibinfo{volume}{59}},
  \bibinfo{pages}{5678} (\bibinfo{year}{1999}).

\bibitem[{\citenamefont{Yu and Cardona}(2005)}]{Peter2005}
\bibinfo{author}{\bibfnamefont{P.~Y.} \bibnamefont{Yu}} \bibnamefont{and}
  \bibinfo{author}{\bibfnamefont{M.}~\bibnamefont{Cardona}},
  \bibinfo{title}{Fundamentals of Semiconductors Physics and Materials
  Properties} (\bibinfo{publisher}{Springer}, \bibinfo{year}{2005}).

\bibitem[{\citenamefont{Williamson et~al.}(2000)\citenamefont{Williamson, Wang,
  and Zunger}}]{WangLW_PRB2000}
\bibinfo{author}{\bibfnamefont{A.~J.} \bibnamefont{Williamson}},
  \bibinfo{author}{\bibfnamefont{L.~W.} \bibnamefont{Wang}}, \bibnamefont{and}
  \bibinfo{author}{\bibfnamefont{A.}~\bibnamefont{Zunger}},
  \bibinfo{journal}{Phys. Rev. B} \textbf{\bibinfo{volume}{62}},
  \bibinfo{pages}{12963} (\bibinfo{year}{2000}).

\bibitem[{\citenamefont{Pryor et~al.}(1998)\citenamefont{Pryor, Kim, Wang,
  Williamson, and Zunger}}]{Zunger_JAP1998}
\bibinfo{author}{\bibfnamefont{C.}~\bibnamefont{Pryor}},
  \bibinfo{author}{\bibfnamefont{J.}~\bibnamefont{Kim}},
  \bibinfo{author}{\bibfnamefont{L.~W.} \bibnamefont{Wang}},
  \bibinfo{author}{\bibfnamefont{A.~J.} \bibnamefont{Williamson}},
  \bibnamefont{and} \bibinfo{author}{\bibfnamefont{A.}~\bibnamefont{Zunger}},
  \bibinfo{journal}{J. Appl. Phys.} \textbf{\bibinfo{volume}{83}},
  \bibinfo{pages}{2548} (\bibinfo{year}{1998}).

\bibitem[{\citenamefont{Luo et~al.}(2005)\citenamefont{Luo, Li, Xia, and
  Wang}}]{Luo_PRB2005}
\bibinfo{author}{\bibfnamefont{J.-W.} \bibnamefont{Luo}},
  \bibinfo{author}{\bibfnamefont{S.-S.} \bibnamefont{Li}},
  \bibinfo{author}{\bibfnamefont{J.-B.} \bibnamefont{Xia}}, \bibnamefont{and}
  \bibinfo{author}{\bibfnamefont{L.-W.} \bibnamefont{Wang}},
  \bibinfo{journal}{Phys. Rev. B} \textbf{\bibinfo{volume}{71}},
  \bibinfo{pages}{245315} (\bibinfo{year}{2005}).

\bibitem[{\citenamefont{Luo et~al.}(2015)\citenamefont{Luo, Bester, and
  Zunger}}]{Luo_PRB2015}
\bibinfo{author}{\bibfnamefont{J.-W.} \bibnamefont{Luo}},
  \bibinfo{author}{\bibfnamefont{G.}~\bibnamefont{Bester}}, \bibnamefont{and}
  \bibinfo{author}{\bibfnamefont{A.}~\bibnamefont{Zunger}},
  \bibinfo{journal}{Phys. Rev. B} \textbf{\bibinfo{volume}{92}},
  \bibinfo{pages}{165301} (\bibinfo{year}{2015}).

\bibitem[{\citenamefont{d'Avezac et~al.}(2012)\citenamefont{d'Avezac, Luo,
  Chanier, and Zunger}}]{Luo_PRL2012}
\bibinfo{author}{\bibfnamefont{M.}~\bibnamefont{d'Avezac}},
  \bibinfo{author}{\bibfnamefont{J.-W.} \bibnamefont{Luo}},
  \bibinfo{author}{\bibfnamefont{T.}~\bibnamefont{Chanier}}, \bibnamefont{and}
  \bibinfo{author}{\bibfnamefont{A.}~\bibnamefont{Zunger}},
  \bibinfo{journal}{Phys. Rev. Lett.} \textbf{\bibinfo{volume}{108}},
  \bibinfo{pages}{027401} (\bibinfo{year}{2012}).

\bibitem[{\citenamefont{Zhang et~al.}(2012)\citenamefont{Zhang, d'Avezac, Luo,
  and Zunger}}]{Luo_NanoLett2012}
\bibinfo{author}{\bibfnamefont{L.-J.} \bibnamefont{Zhang}},
  \bibinfo{author}{\bibfnamefont{M.}~\bibnamefont{d'Avezac}},
  \bibinfo{author}{\bibfnamefont{J.-W.} \bibnamefont{Luo}}, \bibnamefont{and}
  \bibinfo{author}{\bibfnamefont{A.}~\bibnamefont{Zunger}},
  \bibinfo{journal}{Nano Lett.} \textbf{\bibinfo{volume}{12}},
  \bibinfo{pages}{984} (\bibinfo{year}{2012}).

\bibitem[{\citenamefont{Zhang et~al.}(2013)\citenamefont{Zhang, Luo, Saraiva,
  Koiller, and Zunger}}]{Luo_NC2013}
\bibinfo{author}{\bibfnamefont{L.-J.} \bibnamefont{Zhang}},
  \bibinfo{author}{\bibfnamefont{J.-W.} \bibnamefont{Luo}},
  \bibinfo{author}{\bibfnamefont{A.}~\bibnamefont{Saraiva}},
  \bibinfo{author}{\bibfnamefont{B.}~\bibnamefont{Koiller}}, \bibnamefont{and}
  \bibinfo{author}{\bibfnamefont{A.}~\bibnamefont{Zunger}},
  \bibinfo{journal}{Nat. Commun.} \textbf{\bibinfo{volume}{4}},
  \bibinfo{pages}{2396} (\bibinfo{year}{2013}).

\bibitem[{\citenamefont{Wang and Zunger}(1994{\natexlab{b}})}]{Wang_JCP1994}
\bibinfo{author}{\bibfnamefont{L.-W.} \bibnamefont{Wang}} \bibnamefont{and}
  \bibinfo{author}{\bibfnamefont{A.}~\bibnamefont{Zunger}},
  \bibinfo{journal}{J. Chem. Phys.} \textbf{\bibinfo{volume}{100}},
  \bibinfo{pages}{2394} (\bibinfo{year}{1994}{\natexlab{b}}).

\bibitem[{\citenamefont{Luo et~al.}(2017{\natexlab{a}})\citenamefont{Luo, Li,
  and Zunger}}]{Luo2017}
\bibinfo{author}{\bibfnamefont{J.-W.} \bibnamefont{Luo}},
  \bibinfo{author}{\bibfnamefont{S.-S.} \bibnamefont{Li}}, \bibnamefont{and}
  \bibinfo{author}{\bibfnamefont{A.}~\bibnamefont{Zunger}},
  \bibinfo{journal}{Phys. Rev. Lett.} \textbf{\bibinfo{volume}{119}},
  \bibinfo{pages}{126401} (\bibinfo{year}{2017}{\natexlab{a}}).

\bibitem[{\citenamefont{Luo et~al.}(2009)\citenamefont{Luo, Bester, and
  Zunger}}]{Luo_PRL2009}
\bibinfo{author}{\bibfnamefont{J.-W.} \bibnamefont{Luo}},
  \bibinfo{author}{\bibfnamefont{G.}~\bibnamefont{Bester}}, \bibnamefont{and}
  \bibinfo{author}{\bibfnamefont{A.}~\bibnamefont{Zunger}},
  \bibinfo{journal}{Phys. Rev. Lett.} \textbf{\bibinfo{volume}{102}},
  \bibinfo{pages}{056405} (\bibinfo{year}{2009}).

\bibitem[{\citenamefont{Luo et~al.}(2010)\citenamefont{Luo, Chantis, van
  Schilfgaarde, Bester, and Zunger}}]{Luo2010}
\bibinfo{author}{\bibfnamefont{J.-W.} \bibnamefont{Luo}},
  \bibinfo{author}{\bibfnamefont{A.~N.} \bibnamefont{Chantis}},
  \bibinfo{author}{\bibfnamefont{M.}~\bibnamefont{van Schilfgaarde}},
  \bibinfo{author}{\bibfnamefont{G.}~\bibnamefont{Bester}}, \bibnamefont{and}
  \bibinfo{author}{\bibfnamefont{A.}~\bibnamefont{Zunger}},
  \bibinfo{journal}{Phys. Rev. Lett.} \textbf{\bibinfo{volume}{104}},
  \bibinfo{pages}{066405} (\bibinfo{year}{2010}).

\bibitem[{\citenamefont{Luo and Zunger}(2010)}]{LuoJW_PRL2010}
\bibinfo{author}{\bibfnamefont{J.-W.} \bibnamefont{Luo}} \bibnamefont{and}
  \bibinfo{author}{\bibfnamefont{A.}~\bibnamefont{Zunger}},
  \bibinfo{journal}{Phys. Rev. Lett.} \textbf{\bibinfo{volume}{105}},
  \bibinfo{pages}{176805} (\bibinfo{year}{2010}).

\bibitem[{\citenamefont{Luo et~al.}(2017{\natexlab{b}})\citenamefont{Luo, Li,
  Sychugov, Pevere, Linnros, and Zunger}}]{Luo_NatNano2017}
\bibinfo{author}{\bibfnamefont{J.-W.} \bibnamefont{Luo}},
  \bibinfo{author}{\bibfnamefont{S.-S.} \bibnamefont{Li}},
  \bibinfo{author}{\bibfnamefont{I.}~\bibnamefont{Sychugov}},
  \bibinfo{author}{\bibfnamefont{F.}~\bibnamefont{Pevere}},
  \bibinfo{author}{\bibfnamefont{J.}~\bibnamefont{Linnros}}, \bibnamefont{and}
  \bibinfo{author}{\bibfnamefont{A.}~\bibnamefont{Zunger}},
  \bibinfo{journal}{Nat. Nanotech.} \textbf{\bibinfo{volume}{12}},
  \bibinfo{pages}{930} (\bibinfo{year}{2017}{\natexlab{b}}).

\bibitem[{\citenamefont{Rashba and Tela}(1960)}]{Rashba1960}
\bibinfo{author}{\bibfnamefont{E.~I.} \bibnamefont{Rashba}} \bibnamefont{and}
  \bibinfo{author}{\bibfnamefont{F.~T.} \bibnamefont{Tela}},
  \bibinfo{journal}{Phys. Solid State} \textbf{\bibinfo{volume}{2}},
  \bibinfo{pages}{1109} (\bibinfo{year}{1960}).

\bibitem[{\citenamefont{Kloeffel et~al.}(2011)\citenamefont{Kloeffel, Trif, and
  Loss}}]{Kloeffel2011}
\bibinfo{author}{\bibfnamefont{C.}~\bibnamefont{Kloeffel}},
  \bibinfo{author}{\bibfnamefont{M.}~\bibnamefont{Trif}}, \bibnamefont{and}
  \bibinfo{author}{\bibfnamefont{D.}~\bibnamefont{Loss}},
  \bibinfo{journal}{Phys. Rev. B} \textbf{\bibinfo{volume}{84}},
  \bibinfo{pages}{195314} (\bibinfo{year}{2011}).

\bibitem[{\citenamefont{Winkler}(2000)}]{Winkler2000}
\bibinfo{author}{\bibfnamefont{R.}~\bibnamefont{Winkler}},
  \bibinfo{journal}{Phys. Rev. B} \textbf{\bibinfo{volume}{62}},
  \bibinfo{pages}{4245} (\bibinfo{year}{2000}).

\bibitem[{\citenamefont{Marcellina et~al.}(2017)\citenamefont{Marcellina,
  Hamilton, Winkler, and Culcer}}]{Marcellina_PRB2017}
\bibinfo{author}{\bibfnamefont{E.}~\bibnamefont{Marcellina}},
  \bibinfo{author}{\bibfnamefont{A.~R.} \bibnamefont{Hamilton}},
  \bibinfo{author}{\bibfnamefont{R.}~\bibnamefont{Winkler}}, \bibnamefont{and}
  \bibinfo{author}{\bibfnamefont{D.}~\bibnamefont{Culcer}},
  \bibinfo{journal}{Phys. Rev. B} \textbf{\bibinfo{volume}{95}},
  \bibinfo{pages}{075305} (\bibinfo{year}{2017}).

\bibitem[{\citenamefont{Kloeffel et~al.}(2013)\citenamefont{Kloeffel, Trif,
  Stano, and Loss}}]{Kloeffel_PRB2013}
\bibinfo{author}{\bibfnamefont{C.}~\bibnamefont{Kloeffel}},
  \bibinfo{author}{\bibfnamefont{M.}~\bibnamefont{Trif}},
  \bibinfo{author}{\bibfnamefont{P.}~\bibnamefont{Stano}}, \bibnamefont{and}
  \bibinfo{author}{\bibfnamefont{D.}~\bibnamefont{Loss}},
  \bibinfo{journal}{Phys. Rev. B} \textbf{\bibinfo{volume}{88}},
  \bibinfo{pages}{241405(R)} (\bibinfo{year}{2013}).

\bibitem[{\citenamefont{Altmann and Herzig}(2011)}]{Altmann_PG}
\bibinfo{author}{\bibfnamefont{S.~L.} \bibnamefont{Altmann}} \bibnamefont{and}
  \bibinfo{author}{\bibfnamefont{P.}~\bibnamefont{Herzig}},
  \bibinfo{title}{Point group theory tables} (\bibinfo{year}{2011}).

\bibitem[{\citenamefont{Satpathy et~al.}(1988)\citenamefont{Satpathy, Martin,
  and de~Walle}}]{Satpathy_PRB1988}
\bibinfo{author}{\bibfnamefont{S.}~\bibnamefont{Satpathy}},
  \bibinfo{author}{\bibfnamefont{R.~M.} \bibnamefont{Martin}},
  \bibnamefont{and} \bibinfo{author}{\bibfnamefont{C.~G.~V.}
  \bibnamefont{de~Walle}}, \bibinfo{journal}{Phys. Rev. B}
  \textbf{\bibinfo{volume}{38}}, \bibinfo{pages}{13237} (\bibinfo{year}{1988}).

\bibitem[{\citenamefont{Ivchenko et~al.}(1996)\citenamefont{Ivchenko, Kaminski,
  and R\"ossler}}]{Ivchenko1996}
\bibinfo{author}{\bibfnamefont{E.~L.} \bibnamefont{Ivchenko}},
  \bibinfo{author}{\bibfnamefont{A.~Y.} \bibnamefont{Kaminski}},
  \bibnamefont{and}
  \bibinfo{author}{\bibfnamefont{U.}~\bibnamefont{R\"ossler}},
  \bibinfo{journal}{Phys. Rev. B} \textbf{\bibinfo{volume}{54}},
  \bibinfo{pages}{5852} (\bibinfo{year}{1996}).

\bibitem[{\citenamefont{Golub and Ivchenko}(2004)}]{Golub2004}
\bibinfo{author}{\bibfnamefont{L.~E.} \bibnamefont{Golub}} \bibnamefont{and}
  \bibinfo{author}{\bibfnamefont{E.~L.} \bibnamefont{Ivchenko}},
  \bibinfo{journal}{Phys. Rev. B} \textbf{\bibinfo{volume}{69}},
  \bibinfo{pages}{115333} (\bibinfo{year}{2004}).

\bibitem[{\citenamefont{Cartoixa et~al.}(2006)\citenamefont{Cartoixa, Wang,
  Ting, and Chang}}]{Cartoixa2006}
\bibinfo{author}{\bibfnamefont{X.}~\bibnamefont{Cartoixa}},
  \bibinfo{author}{\bibfnamefont{L.-W.} \bibnamefont{Wang}},
  \bibinfo{author}{\bibfnamefont{D.~Z.-Y.} \bibnamefont{Ting}},
  \bibnamefont{and} \bibinfo{author}{\bibfnamefont{Y.-C.} \bibnamefont{Chang}},
  \bibinfo{journal}{Phys. Rev. B} \textbf{\bibinfo{volume}{73}},
  \bibinfo{pages}{205341} (\bibinfo{year}{2006}).

\bibitem[{\citenamefont{Nestoklon et~al.}(2008)\citenamefont{Nestoklon,
  Ivchenko, Jancu, and Voisin}}]{Nestoklon2008}
\bibinfo{author}{\bibfnamefont{M.~O.} \bibnamefont{Nestoklon}},
  \bibinfo{author}{\bibfnamefont{E.~L.} \bibnamefont{Ivchenko}},
  \bibinfo{author}{\bibfnamefont{J.~M.} \bibnamefont{Jancu}}, \bibnamefont{and}
  \bibinfo{author}{\bibfnamefont{P.}~\bibnamefont{Voisin}},
  \bibinfo{journal}{Phys. Rev. B} \textbf{\bibinfo{volume}{77}},
  \bibinfo{pages}{155328} (\bibinfo{year}{2008}).

\bibitem[{\citenamefont{Kitaev et~al.}(1997)\citenamefont{Kitaev, Panfilov,
  Tronc, and Evarestov}}]{Kitaev1997}
\bibinfo{author}{\bibfnamefont{Y.~E.} \bibnamefont{Kitaev}},
  \bibinfo{author}{\bibfnamefont{A.~G.} \bibnamefont{Panfilov}},
  \bibinfo{author}{\bibfnamefont{P.}~\bibnamefont{Tronc}}, \bibnamefont{and}
  \bibinfo{author}{\bibfnamefont{R.~A.} \bibnamefont{Evarestov}},
  \bibinfo{journal}{Journal of Physics: Condensed Matter}
  \textbf{\bibinfo{volume}{9}}, \bibinfo{pages}{257} (\bibinfo{year}{1997}),
  ISSN \bibinfo{issn}{1361-648X}.

\bibitem[{\citenamefont{Cheng et~al.}(2021)\citenamefont{Cheng, Kesan,
  Grutzmacher, Sedgwick, and Ott}}]{Cheng2021}
\bibinfo{author}{\bibfnamefont{J.}~\bibnamefont{Cheng}},
  \bibinfo{author}{\bibfnamefont{V.~P.} \bibnamefont{Kesan}},
  \bibinfo{author}{\bibfnamefont{D.~A.} \bibnamefont{Grutzmacher}},
  \bibinfo{author}{\bibfnamefont{T.~O.} \bibnamefont{Sedgwick}},
  \bibnamefont{and} \bibinfo{author}{\bibfnamefont{J.~A.} \bibnamefont{Ott}},
  \bibinfo{journal}{Appl. Phys. Lett.} \textbf{\bibinfo{volume}{62}},
  \bibinfo{pages}{1522} (\bibinfo{year}{2021}).

\bibitem[{\citenamefont{Schuberth et~al.}(1991)\citenamefont{Schuberth,
  Abstreiter, Gornik, Sch\"affler, and Luy}}]{Schuberth1991}
\bibinfo{author}{\bibfnamefont{G.}~\bibnamefont{Schuberth}},
  \bibinfo{author}{\bibfnamefont{G.}~\bibnamefont{Abstreiter}},
  \bibinfo{author}{\bibfnamefont{E.}~\bibnamefont{Gornik}},
  \bibinfo{author}{\bibfnamefont{F.}~\bibnamefont{Sch\"affler}},
  \bibnamefont{and} \bibinfo{author}{\bibfnamefont{J.~F.} \bibnamefont{Luy}},
  \bibinfo{journal}{Phys. Rev. B} \textbf{\bibinfo{volume}{43}},
  \bibinfo{pages}{2280} (\bibinfo{year}{1991}).

\bibitem[{\citenamefont{Belhadj et~al.}(2021)\citenamefont{Belhadj, Amand,
  Kunold, Simon, Kuroda, Abbarchi, Mano, Sakoda, Kunz, Marie
  et~al.}}]{Belhadj2021}
\bibinfo{author}{\bibfnamefont{T.}~\bibnamefont{Belhadj}},
  \bibinfo{author}{\bibfnamefont{T.}~\bibnamefont{Amand}},
  \bibinfo{author}{\bibfnamefont{A.}~\bibnamefont{Kunold}},
  \bibinfo{author}{\bibfnamefont{C.-M.} \bibnamefont{Simon}},
  \bibinfo{author}{\bibfnamefont{T.}~\bibnamefont{Kuroda}},
  \bibinfo{author}{\bibfnamefont{M.}~\bibnamefont{Abbarchi}},
  \bibinfo{author}{\bibfnamefont{T.}~\bibnamefont{Mano}},
  \bibinfo{author}{\bibfnamefont{K.}~\bibnamefont{Sakoda}},
  \bibinfo{author}{\bibfnamefont{S.}~\bibnamefont{Kunz}},
  \bibinfo{author}{\bibfnamefont{X.}~\bibnamefont{Marie}},
  \bibnamefont{et~al.}, \bibinfo{journal}{Appl. Phys. Lett.}
  \textbf{\bibinfo{volume}{97}}, \bibinfo{pages}{051111}
  (\bibinfo{year}{2021}).

\bibitem[{\citenamefont{Hubers et~al.}(2005)\citenamefont{Hubers, Pavlov, and
  Shastin}}]{Huebers2005}
\bibinfo{author}{\bibfnamefont{H.-W.} \bibnamefont{Hubers}},
  \bibinfo{author}{\bibfnamefont{S.~G.} \bibnamefont{Pavlov}},
  \bibnamefont{and} \bibinfo{author}{\bibfnamefont{V.~N.}
  \bibnamefont{Shastin}}, \bibinfo{journal}{Semiconductor Science and
  Technology} \textbf{\bibinfo{volume}{20}}, \bibinfo{pages}{S211}
  (\bibinfo{year}{2005}).

\bibitem[{\citenamefont{Durnev et~al.}(2014)\citenamefont{Durnev, Glazov, and
  Ivchenko}}]{Durnev_PRB2014}
\bibinfo{author}{\bibfnamefont{M.~V.} \bibnamefont{Durnev}},
  \bibinfo{author}{\bibfnamefont{M.~M.} \bibnamefont{Glazov}},
  \bibnamefont{and} \bibinfo{author}{\bibfnamefont{E.~L.}
  \bibnamefont{Ivchenko}}, \bibinfo{journal}{Phys. Rev. B}
  \textbf{\bibinfo{volume}{89}}, \bibinfo{pages}{075430}
  (\bibinfo{year}{2014}).

\bibitem[{\citenamefont{Luttinger}(1956)}]{Luttinger1956}
\bibinfo{author}{\bibfnamefont{J.~M.} \bibnamefont{Luttinger}},
  \bibinfo{journal}{Phys. Rev.} \textbf{\bibinfo{volume}{102}},
  \bibinfo{pages}{1030} (\bibinfo{year}{1956}).

\end{thebibliography}

\end{document}